\documentclass[12pt]{article}

\usepackage[a4paper,margin=3cm]{geometry}
\usepackage[T1]{fontenc}

\usepackage{authblk}

\usepackage{sectsty}
\sectionfont{\fontsize{16}{20}\selectfont}
\subsectionfont{\fontsize{14}{18}\normalfont\em\selectfont}

\usepackage{titlesec}
\titlelabel{\thetitle.\,}

\usepackage[parfill]{parskip}
\usepackage{setspace}

\usepackage[labelfont=bf,font={small,stretch=0.95}]{caption}

\usepackage[square,super,sort,compress,numbers]{natbib}
\usepackage{hyperref}

\usepackage{gensymb}
\usepackage{amsmath}
\usepackage{amssymb}

\usepackage{graphicx}

\usepackage{booktabs}

\usepackage[version=4]{mhchem}

\usepackage{mathptmx}

\begin{document}

\title{\fontsize{20}{26}\sffamily\bf\selectfont The Zintl--Klemm Concept in the Amorphous State: A Case Study of Na--P Battery Anodes}

\author[1]{Litong Wu}
\author[1]{Volker~L.~Deringer\thanks{volker.deringer@chem.ox.ac.uk}}

\affil[1]{Inorganic Chemistry Laboratory, Department of Chemistry, University of Oxford, \protect\\ Oxford OX1 3QR, United Kingdom}

\date{}
\maketitle
\thispagestyle{empty}

\begin{abstract}
\normalsize
The Zintl--Klemm concept has long been used to explain and predict the bonding, and thereby the structures, of crystalline solid-state materials. We apply this concept to the amorphous state, examining as an example the diverse disordered Na--P phases that can form in sodium-ion battery anodes. Using first-principles simulations combined with state-of-the-art machine-learning methods, we provide atomic-scale insight into the structural and energetic behaviour of amorphous Na--P phases. We evaluate the applicability of the Zintl--Klemm rules in the amorphous state and discuss implications for future work.
\end{abstract}

\clearpage 

\setstretch{1.5}

Since its early foundation in the~1930s, the Zintl--Klemm concept has remained a powerful framework for describing the bonding and structure of compounds formed between electropositive alkali and alkaline-earth metals, and electronegative elements from groups~13 to~16 of the Periodic Table.~\cite{Nesper-2014, Kauzlarich-2023} In this concept, the alkali or alkaline-earth metals formally donate their valence electrons to the more electronegative $p$-block elements, which in turn achieve closed-shell electronic configurations by accepting the transferred electrons and forming covalent bonds among themselves.~\cite{Kauzlarich-2016} As a result, Zintl phases exhibit salt-like characteristics due to the ionic interactions between cations and polyanionic units.~\cite{Atkins-2010} 
In the 1950s, Klemm further built on Zintl's idea by introducing \textit{the pseudoatom model}, in which the anionic units are treated as elements with the same number of valence electrons.~\cite{Klemm-1958, Schafer-1973} 
Klemm's work set the stage for iso- and aliovalent modifications in materials design, enabling the guided synthesis of pseudoatom-inspired compounds.~\cite{Nesper-2014, Kauzlarich-2016}

Zintl phases span the spectrum between classical salts and intermetallic compounds, encompassing materials with a diverse range of structural and electronic properties.~\cite{Toberer-2010, Kauzlarich-2023} Various subclasses, including both ordered crystalline phases and those exhibiting correlated disorder, have been extensively studied for different applications. For instance, structurally disordered Zintl phases have been widely investigated in the context of thermoelectric materials, where the presence of defects and disordered substructures can be crucial for tuning lattice thermal conductivity and electronic transport.~\cite{Zeier-2016, Chen-2021} Despite their lack of long-range order, these systems typically yield diffraction patterns with well-defined Bragg peaks.~\cite{Chen-2019, Roth-2021} Fully amorphous Zintl phases remain largely unexplored, with only occasional reports and few systematic studies to date.~\cite{Nesper-2014}

An area in which Zintl-phase amorphization is frequently observed is alloy-based battery anodes, typically group-14 (Si,~\cite{Loaiza-2020} Ge,~\cite{Loaiza-2020} Sn~\cite{Baggetto-2013, Ellis-2012}) or group-15 (P, Sb~\cite{Baggetto-2013-06, Darwiche-2012}) elements, which store Li,~Na, or~K ions through binary alloying or conversion reactions. Unlike durable inter\-calation-based electrode systems, these materials undergo repeated structural destruction and reconstruction during metal insertion and extraction.~\cite{Yabuuchi-2014} While offering high theoretical capacities and low discharge voltages, their practical application is often limited by substantial volume expansion and unstable interfaces.~\cite{Choi-2016} Understanding the mechanisms during the alloy-based anode operation is therefore crucial, especially for Na-ion batteries~(SIBs) which are considered a more cost-effective and sustainable alternative to Li-ion batteries, but still face challenges in anode design. In particular, the larger ionic radius of Na prevents efficient intercalation into commercial graphite anodes, necessitating the search for alternative high-capacity anode materials.~\cite{Yabuuchi-2014}

Here, we focus on phosphorus, a promising SIB anode material with a theoretical capacity of~2,596~mA~h~g$^{-1}$ for its fully sodiated phase, \ce{Na3P}.~\cite{Kim-2013} Both crystalline black and amorphous red P have been extensively investigated experimentally as SIB anode materials. Black P is a semiconducting van-der-Waals material with puckered six-membered rings within its layers.~\cite{Brown-1965, Kolesnik-Gray-2023} The commercially available red P, in contrast, comprises diverse cluster fragments primarily formed from interconnected five- and six-membered rings.~\cite{Zhou-2023} These local structural motifs closely resemble those found in violet and fibrous P~\cite{Elliott-1985, Ruck-2005} as well as P nanorods.~\cite{Pfitzner-2004, Bachhuber-2014} Bulk and monolayer black P~\cite{Sun-2015, Xu-2016, Marbella-2018} as well as bulk red P~\cite{Kim-2013, Qian-2013, Ramireddy-2015, Capone-2020} have been integrated with conductive carbon materials, forming nanostructured composites that showed excellent electrochemical performance in SIBs. While carbon materials contribute minimally to the capacity, it serves as a mechanical backbone and an electron-conducting matrix, enhancing cycling stability. Notably, most studies have identified amorphous Na--P~(a-Na$_x$P) phases as key intermediates during battery cycling.~\cite{Kim-2013, Sun-2015, Xu-2016, Marbella-2018}

To address the challenge of understanding these amorphous phases at the atomic scale, machine-learning methods provide a powerful tool by enabling efficient first-principles atomistic simulations.~\cite{Behler-2017, Deringer-2019, Friederich-2021} Machine-learning-based interatomic potential (MLIP) models have been applied to elemental P to describe the phase transition between molecular and network liquid forms,~\cite{Deringer-2020} explore hypothetical hierarchi\-cal\-ly-structured allotropes,~\cite{Deringer-2020a} and study the structural and bonding characteristics of amorphous red P.~\cite{Zhou-2022, Zhou-2023} Building on these advances, our present work moves beyond the elemental system and explores the structural and energetic landscape of a-Na$_x$P compounds. Our simulations are based on a custom-fitted MLIP model developed using the MACE architecture,\cite{Batatia-2023} which achieves root-mean-square errors of 7.7~meV~at.$^{-1}$ for energy and 0.13~eV~\r{A}$^{-1}$ for force predictions (Figure~S2). Full details of the computational approach, including dataset construction and training protocols, are provided in the Supporting Information.

\begin{figure}[t!]
    \centering
    \includegraphics[width=\linewidth]{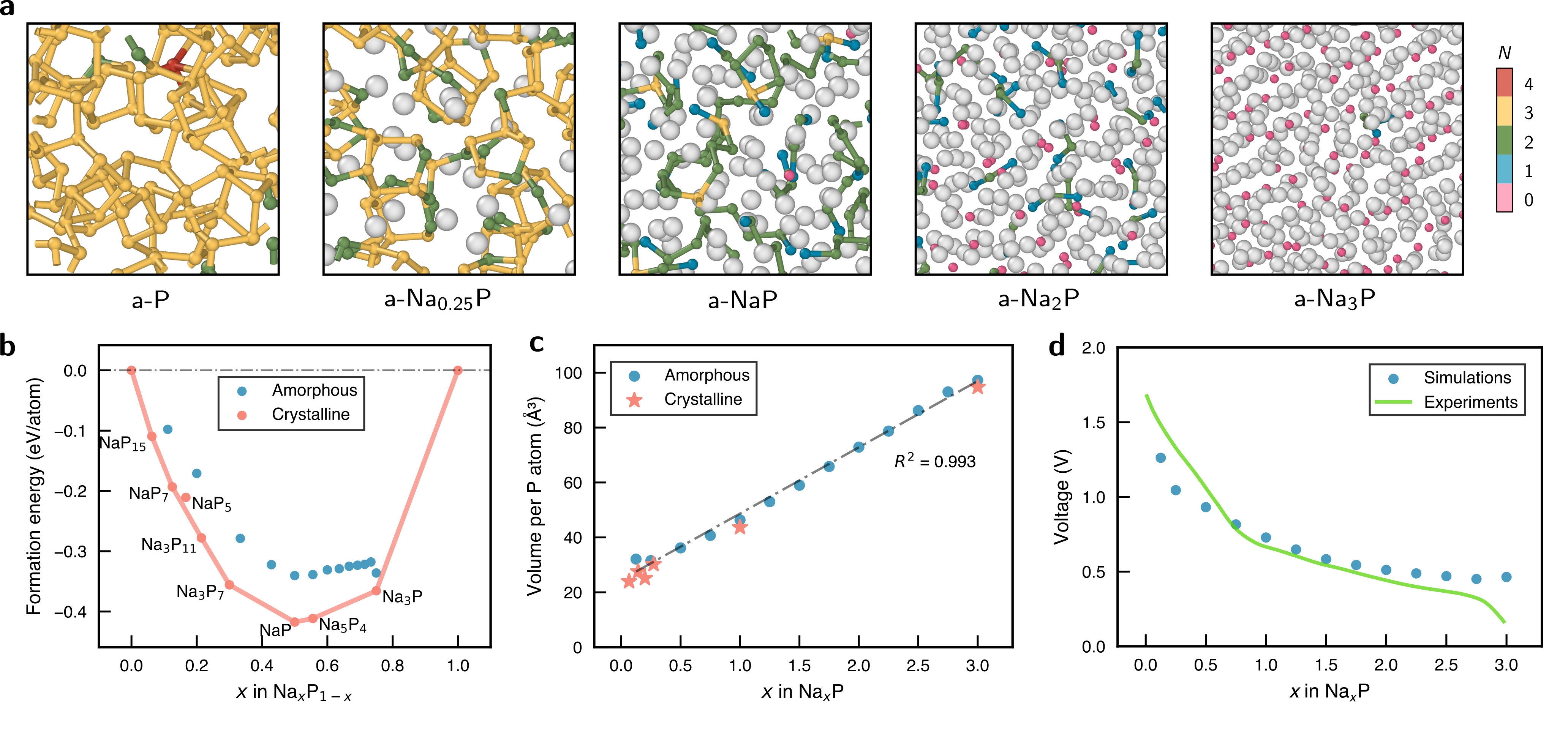}
    \caption{Amorphous Na--P phases from ML-driven simulations. (\textbf{a}) Representative structures of a-P and a-Na{$_x$}P obtained from melt--quench simulations, shown in order of increasing Na content (grey spheres). P atoms are colour-coded by homoatomic coordination number: $\textit{N} = \text{4}$~(red), $\textit{N} = \text{3}$~(yellow), $\textit{N} = \text{2}$~(green), $\textit{N} = \text{1}$~(blue), and $\textit{N} = \text{0}$~(pink) P neighbours, determined using a 2.4~\r{A} cutoff. The cell images have been cropped for visualization, and hence the borders do not represent the simulation cell edges. (\textbf{b}) Normalized formation energies as a function of fractional Na content in Na--P compounds, referenced to black P and body-centred cubic Na. The convex hull is constructed by joining stable crystalline Na--P phases (red); amorphous phases are shown in blue. MACE energies are used for amorphous structures; DFT energies for crystalline structures. (\textbf{c}) Simulation cell volumes normalized by the number of P atoms, plotted against Na content. A best-fit line for the amorphous-phase volumes is shown. (\textbf{d}) Voltage profile for the sodiation of amorphous P. Experimental data from Capone~et~al. (Ref.~\citenum{Capone-2020}) are shown by a green line.}
    \label{fig:1}
\end{figure}

To systematically explore the compositional range from elemental P to \ce{Na3P}, we performed ML-driven melt--quench mole\-cular-dynamics (MD) simulations to generate a series of a-Na$_x$P structures --- each containing 248~P atoms with varying Na content, up to a maximum of 744~Na atoms ($x$~=~3). Initially, pristine a-P structures were generated following established protocols.~\cite{Zhou-2022, Zhou-2023} To accommodate Na atoms within the covalent P frameworks, it was necessary to expand the a-P cells: target volumes per P atom for different Na content were estimated via linear least-squares regression of crystalline-phase volumes (Supporting Information). After isotropically expanding the cells to the calculated volumes, Na atoms were inserted at random positions with a hard-sphere cutoff. The resulting structures were melted at 1,200~K and quenched at a rate of 10$^{12}$~K~s$^{-1}$ to obtain amorphous structure models. Selected a-P and a-Na$_x$P structures are shown in Figure~\ref{fig:1}a, using smaller 108-P systems for visual clarity. Na atoms are shown as grey spheres, while P atoms are colour-coded according to their homonuclear coordination number, $N$. From left to right, as the Na content increases, the P framework progressively breaks down, transitioning from an extended network to branched chains, chain fragments, and eventually isolated ions.

To assess the thermodynamic stability of the amorphous phases, we constructed the convex hull using the formation energies of stable crystalline Na--P phases (Figure 1b), calculated from DFT energies akin to Ref.~\citenum{Mayo-2016}. All on-hull phases have been experimentally observed, with the exception of \ce{Na5P4}, which was theoretically predicted to be stable.~\cite{Mayo-2016} The known \ce{NaP5} phase,~\cite{Chen-2004} although located above the computed convex hull, is included for comparison. Formation energies of the amorphous samples, calculated using our MACE model, are plotted relative to the convex hull. As expected, the a-Na$_x$P phases lie above the hull, reflecting their lower thermodynamic stability compared to crystalline structures. Notably, our a-\ce{Na3P} sample exhibits an anomalous stabilization, likely suggesting a tendency to crystallize.

Since structural expansion is a key challenge that limits the longevity of P anodes, simulation cell volumes normalized by the number of P atoms are plotted in Figure~\ref{fig:1}c. The data reveal an almost linear increase in normalized volume as the number of Na atoms increases, with a strong correlation ($R^2 = 0.993$). The volumes of the amorphous phases are comparable to those of their crystalline counterparts, and including both crystalline and amorphous structures in a linear regression reduces the $R^2$ value by less than 0.001. In particular, the a-\ce{Na3P} structure expands by 329\% relative to the initial a-P structure, closely matching the theoretical prediction of 331\%.\cite{Capone-2020}

Another experimentally relevant quantity, the voltage at different stages of sodiation, was estimated computationally similar to previous work~\cite{Huang-2019} (Figure~\ref{fig:1}d). An experimental charging profile of Na with ball-milled red-P--graphite electrodes, from Ref.~\citenum{Capone-2020}, is plotted in green for comparison. A reasonably good agreement is seen in the intermediate composition range, supporting the validity of the MLIP model. However, it is important to note that experimentally measured voltages can be affected by side reactions and kinetic limitations, not captured on the time scale of typical MD simulations. The selected experimental data, based on steady-state potentials from Galvanostatic Intermittent Titration Technique measurements,~\cite{Zhu-2010} offer a more reliable estimate of the true equilibrium potential across different states of charge.

\begin{figure}
    \centering
    \includegraphics[width=\linewidth]{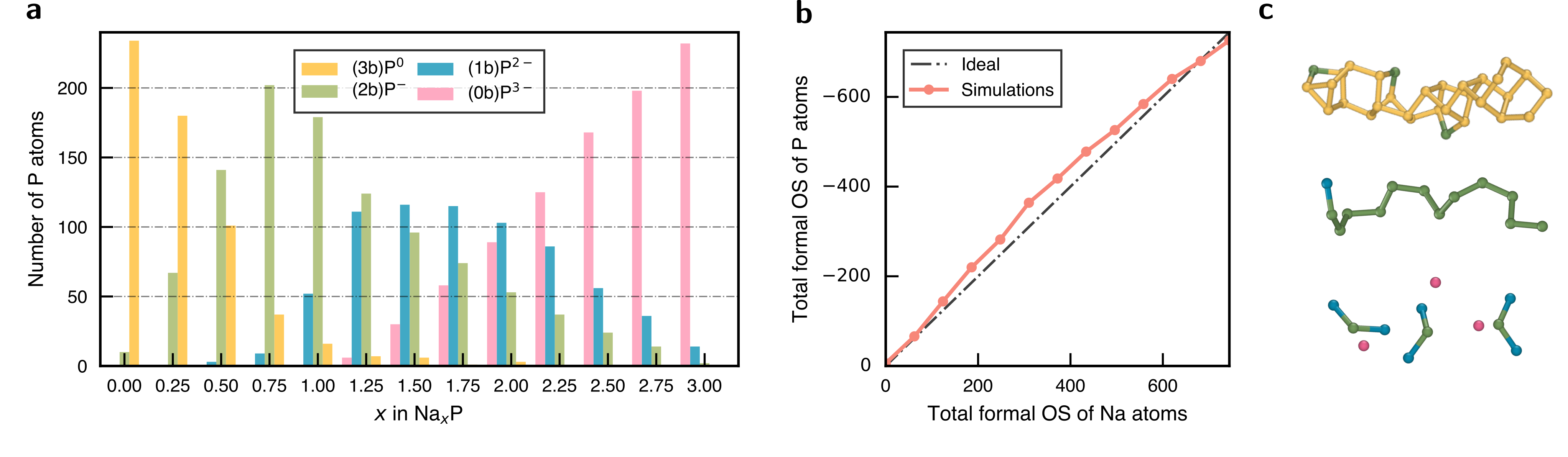}
    \caption{The Zintl--Klemm concept applied to amorphous Na--P phases. (\textbf{a}) Distribution of P atoms with different homoatomic coordination numbers in each a-Na$_x$P configuration across the compositional range. P atoms are classified by a connectivity symbol in parentheses and the formal charge as superscript. (\textbf{b}) Total formal oxidation state (OS) of P atoms compared to Na atoms in each a-Na$_x$P structure. The ideal values are shown by a dotted line. (\textbf{c}) Selected local structural fragments from a-Na$_{\text{0.125}}$P, a-NaP, a-Na$_\text{2}$P (top to bottom).}
    \label{fig:2}
\end{figure}

We next examine the applicability of the Zintl--Klemm concept to a-Na$_x$P structures in Figure~\ref{fig:2}. In this analysis, P atoms in all a-Na$_x$P structures are classified and colour-coded according to their respective homonuclear connectivity; the labels (0b), (1b), (2b), and (3b) denote P atoms with zero, one, two, or three P--P bonds. Formal charges are then assigned as a superscript based on the valence electron requirements. According to the Zintl--Klemm concept, two key implications arise: first, the total formal oxidation state of the P framework should balance out that of the Na atoms in each a-Na$_x$P structure; second, the resulting `pseudoelements' formed by P after the electron transfer may adopt structural motifs analogous to those of their isoelectronic counterparts.

Figure~\ref{fig:2}a shows the distribution of P atoms in each connectivity category across the compositional range in the amorphous structures. As expected, with increasing Na content, greater proportions of P atoms exhibit lower homonuclear connectivity, indicating P--P covalent bond breaking induced by Na--P interactions. This trend aligns with the chemical expectation that electron donation from Na atoms reduces the need for P--P bonding to satisfy valence requirements. Interestingly, the proportion of (1b)P$^{2-}$ species remains comparatively low across all compositions, even in regimes where it would be expected to dominate. In addition, the corresponding total formal oxidation state for each configuration, shown in Figure~\ref{fig:2}b, conforms well to the ideal Zintl--Klemm formulation indicated by a dotted line. This implies that, even in the absence of long-range order, atoms arrange locally to preserve overall charge balance --- reflecting the continued relevance of classical chemical principles in complex disordered systems.

Representative local structural fragments from our simulations are shown in Figure~\ref{fig:2}c. The top fragment, taken from an a-Na$_{0.125}$P cell, exhibits structural features reminiscent of a-P, with clusters composed of interconnected five- and six-membered rings.~\cite{Bocker-1995} Within these clusters, (2b)P$^{-}$ ions are often found near Na atoms --- an arrangement also observed in the mixed-valence polymeric P framework of crystalline NaP$_{15}$~\cite{Grotz-2015} and NaP$_{7}$~\cite{Grotz-2015-01} (Figure~S11). As sodiation proceeds, these clusters gradually break apart, giving rise to chain-like motifs, that dominate near the NaP stoichiometry. Compared to the helical chain found in crystalline NaP,~\cite{Georg-1979} chains in the amorphous structures do not exhibit local symmetry, with occasional branching and termination. This transition in the dominant structural motifs highlights the second key implication of the Zintl--Klemm concept: in a-NaP, the formally (2b)P$^{-}$ ions are isoelectronic to S atoms. Indeed, chain-like motifs are common in the allotropes of S,~\cite{Crichton-2001} Se,~\cite{Geller-1967} and Te.~\cite{Cherin-1967, Keller-1977} 

Upon further sodiation, these chains progressively fragment. In a-Na$_{2}$P, the (1b)P$^{2-}$ ions, isoelectronic to Cl, might be expected to form dumbbells with isolobal analogy to \ce{Cl2} molecules. However, such dumbbell-like motifs are not prevalent in the a-Na$_{2}$P samples (cf.\ Figure~\ref{fig:1}a). Instead, the dominant structural motifs at this composition are bent triatomic molecules, accompanied by isolated (0b)P$^{3-}$ ions, as shown at the bottom of Figure~\ref{fig:2}c. This trimer species, (1b)P$^{2-}$--(2b)P$^{-}$--(1b)P$^{2-}$, is both isoelectronic and isostructural to the well-known SCl$_{2}$ mole\-cule.  Principally, two dumbbell units are electronically equivalent to one bent trimer and one isolated phosphide anion, and the interconversion between these species can theoretically be achieved via a (formal) disproportionation reaction:

\begin{figure}[h!]
    \centering
    \includegraphics[width=0.5\linewidth]{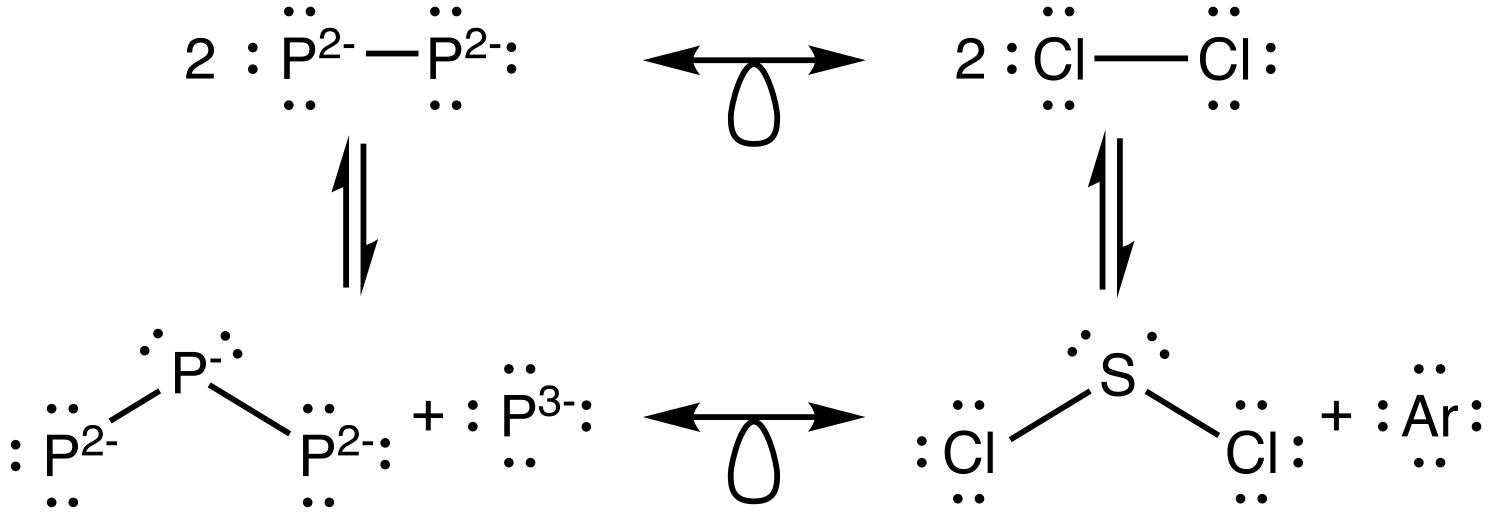}
    \label{fig:enter-label}
\end{figure}

\begin{figure}[t!]
    \centering
    \includegraphics[width=0.5\linewidth]{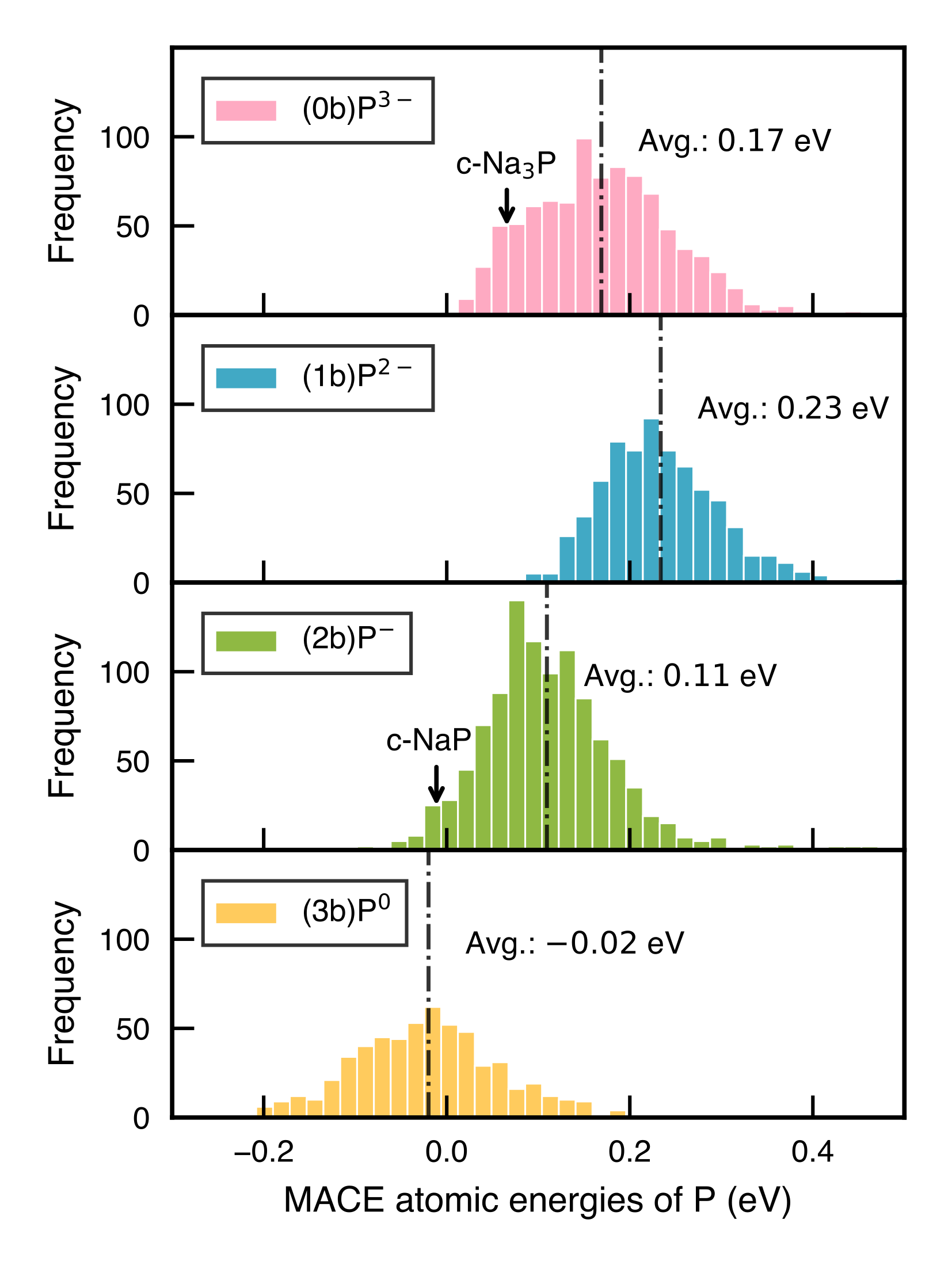}
    \caption{Distributions of the MACE atomic energies of P atoms in a-Na$_x$P ($\text{0.125} \leq x \leq \text{3}$) structures, shown separately for P with different local connectivity. All values are referenced to the atomic energy of crystalline black P. The atomic energies of crystalline Na$_\text{3}$P and NaP are marked with arrows. The average atomic energy for each category is indicated by a vertical dashed line. A similar analysis for Na atoms is discussed in the Supporting Information.}
    \label{fig:3}
\end{figure}

To understand the underrepresentation of (1b)P$^{2-}$ and the preference for \ce{SCl2}-like trimers over \ce{Cl2}-like dumbbells, we analysed the MACE-predicted local {\em per-atom} energy distributions for P atoms with different local connectivity in all a-Na$_x$P structures, as shown in Figure~\ref{fig:3}. Although the interpretability of MLIP atomic energies remains a topic of debate, growing evidence suggests that these local energy values can provide meaningful insights.~\cite{Bernstein-2019, Wang-2023, Chong-2023, Morrow-2024} 
Here, we observe that the average atomic energy of (1b)P$^{2-}$ exceeds that of (0b)P$^{3-}$, and is in fact the highest among all four types of P environments in a-Na$_x$P. This observation may help explain the preferential formation of bent trimers with isolated phosphide anions over dumbbells: despite having the same overall formal oxidation state, the former incurs a lower energetic cost by having two (1b)P$^{2-}$ ions instead of four. The instability of dumbbell-like motifs is also consistent with previous experimental and computational findings. $^{31}$P NMR measurements by Marbella et al.\ suggested that dumbbell-containing structures were not present in significant quantities in SIBs during cycling.~\cite{Marbella-2018} Similarly, \textit{ab initio} studies of Li- and Na-ion anode materials by Mayo et al.\ revealed that while P dumbbells are common in Li$_{x}$P when $1.33 < x \leq 2$, they are absent in the Na--P system.~\cite{Mayo-2016} The same study identified Na$_4$P$_3$, a locally stable structure only 2~meV per formula unit above the convex hull, as comprising bent P trimers similar to those observed in our amorphous structures.

To further understand the local behaviour of P atoms, atomic charges were computed using the Löwdin scheme~\cite{Lowdin-1950} as implemented in LOBSTER.\cite{Maintz-2016, Ertural-2019} An independent set of smaller a-Na$_x$P structures (each containing 108~P atoms) was generated following the same melt--quench protocol to perform the charge calculations. Table~\ref{table:1} summarizes the Löwdin charges of P atoms with different local coordination: as expected, P species with more negative formal oxidation states based on the coordination analysis also exhibit more negative average Löwdin charges. The relatively small standard deviations compared to the differences in mean values suggest that the charge distributions within each coordination class are fairly narrow, indicating a clear differentiation across bonding environments.

\begin{table}[t]
    \centering
    \caption{Average Löwdin charge and corresponding standard deviation for P with different local connectivity in 108-P a-Na$_x$P ($\text{0.25} \leq x \leq \text{3}$) structures. A similar analysis for Na is discussed in the Supporting Information.}
    \begin{tabular}{lcc} 
                \toprule
                \hspace{1em}\textbf{Coordination}\hspace{1em} & \textbf{Charge (\boldmath $e$)} \\ 
                \midrule
                \hspace{2.5em}(0b)P$^{3-}$ & \hspace{1em}$-$1.447 $\pm$ 0.031\hspace*{1em}\\
                \hspace{2.5em}(1b)P$^{2-}$ & \hspace{1em}$-$0.984 $\pm$ 0.060\hspace*{1em}\\
                \hspace{2.5em}(2b)P$^{2-}$ & \hspace{1em}$-$0.470 $\pm$ 0.100\hspace*{1em}\\
                \hspace{2.5em}(3b)P$^{0}$ & \hspace{1em}$-$0.066 $\pm$ 0.040\hspace*{1em}\\
                \bottomrule
    \end{tabular}
    \label{table:1}
\end{table}

We note that the magnitude of Löwdin charges is consistently smaller than that of the formal charges. This difference is to be expected: the Zintl model is a valence-based heuristic assuming full electron transfer, which can lead to overestimated charges; Löwdin charges are derived from quantum-mechanical orbital projections, which are reported to systematically underestimate ionic character in strongly ionic systems.~\cite{Kar-1988} And yet, the charge analysis provides a quantitative complement to the Zintl--Klemm concept --- it confirms the electronic distinction between different P coordination classes, and highlights the increasingly ionic character of P species in a-Na$_x$P structures with increasing Na content.

\begin{figure*}[t!]
    \centering
    \includegraphics[width=\linewidth]{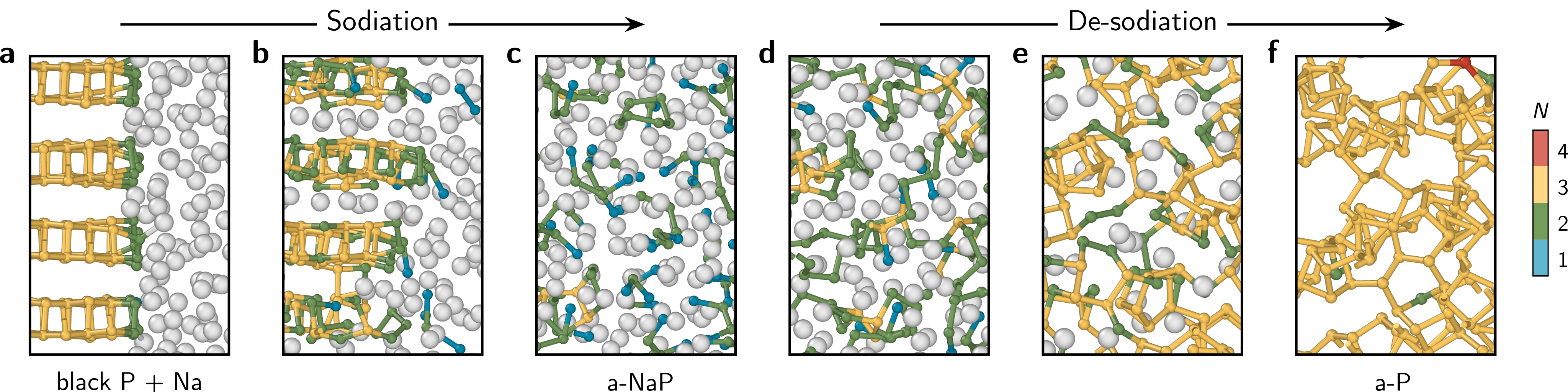}
    \caption{Structural change of black P during sodiation and de-sodiation. (\textbf{a}) Initial configuration of the simulation cell, with a 256-atom black P supercell positioned on the left and Na atoms randomly placed in the adjacent space on the right. (\textbf{b}) Intermediate configuration during sodiation, showing Na intercalation and partial P framework destruction. (\textbf{c}) Final configuration of the 1:1 (Na:P) sodiation simulation. (\textbf{d}--\textbf{f}) Final configurations following the removal of 25\%, 75\%, and 100\% of the Na atoms, respectively. The cell images have been cropped for visualization.}
    \label{fig:4}
\end{figure*}

Finally, a series of small-scale, proof-of-concept MACE-driven MD simulations were carried out to investigate the structural evolution during sodiation and de-sodiation, inspired by earlier \textit{ab initio} MD studies.~\cite{Johari-2011} The initial configuration (Figure~\ref{fig:4}a) was based on a black P supercell containing four layers (256 P atoms), positioned along one side of an elongated simulation cell. An equal number of Na atoms were randomly distributed in the adjacent space along the $x$-direction. This orientation was chosen because only channels along the $x$-axis provide sufficient width (3.08~\r{A}) to accommodate the diffusion of Na ions (diameter $= 2.04$~\r{A}~\cite{Shannon-1976}), whereas channels along the $y$-axis are significantly narrower (1.16~\r{A}) and therefore inaccessible.~\cite{Sun-2015} 

The sodiation simulation involved an NPT annealing at 600~K for 100~ps, followed by a rapid quench at a rate of 10$^{13}$~K~s$^{-1}$. At elevated temperatures, molten Na atoms rapidly diffused into the van-der-Waals gaps between the layers, driven by the thermodynamics of Na--P interaction (Figure~\ref{fig:4}b). This Na infiltration initiated P--P bond rearrangements and breakage, eventually leading to the formation of a homogeneous amorphous Na--P phase (Figure~\ref{fig:4}c).
To simulate the de-sodiation process, Na atoms were systematically removed from the relaxed sodiated structure in four steps. Increments of 25\% of the total Na content were extracted sequentially starting from atoms farthest from the originally P-rich region along the $x$-axis. After each extraction, the system was subjected to annealing at 600~K for 100~ps followed by quenching to allow for structural reorganization. This process was repeated four times until all Na atoms had been removed. The relaxed configurations after the removal of 25\%, 75\%, and 100\% of Na atoms are displayed in Figure~\ref{fig:4}d--f.

During the simulation, progressive de-sodiation led to a gradual reformation of P--P bonds. However, despite the reappearance of local bonding motifs, the original layered structure of black P was not recovered. Instead, the system evolved into an amorphous phase with a disordered network topology. Cluster analysis (Figure~S9) reveals a consistent increase in the number of five-membered rings and local cluster fragments, such as P3]P2[P3 and P2[P3]P2 Baudler units,~\cite{Bocker-1995} throughout the de-sodiation process, while the number of six-membered rings showed only a marginal increase. Since six-membered rings are characteristic of crystalline black P, whereas five-membered rings and local cluster fragments are commonly found in amorphous P,~\cite{Zhou-2022} this observation suggests an irreversible amorphization of the P structure --- in agreement with previous experimental reports of irreversible capacity loss and structural disorder in cycled P anodes.~\cite{Kim-2013, Sun-2015, Marbella-2018}

In conclusion, we have presented an ML-driven approach for modelling and interpreting the complex chemical and structural features of the amorphous Na--P system. Through coordination and charge analysis, we showed how the traditional Zintl--Klemm concept remains applicable for highly disordered materials. Furthermore, our pilot simulations in Figure \ref{fig:4} already captured diffusion-driven structural evolution during (de-) sodiation, providing insight into the irreversible amorphization of P anodes. Extending this approach to related systems, such as Li--P or Na--Sb, could enable comparative insights into the chemistry of alloy-type battery materials more widely. Future work could focus on incorporating transport property characterization, with the long-term goal of guiding the design and laboratory synthesis of improved anode materials.

\clearpage

\setstretch{1.1}

\section*{Acknowledgements}

We thank Prof.\ M.\ Pasta for helpful discussions, and Z.\ El-Machachi and Y.\ Zhou for technical help.
L.W. acknowledges funding from the EPSRC Centre for Doctoral Training in Inorganic Chemistry for Future Manufacturing (OxICFM), EP/S023828/1. 
This work was supported by UK Research and Innovation [grant number EP/X016188/1].
We are grateful for computational support from the UK national high performance computing service, ARCHER2, for which access was obtained via the UKCP consortium and funded by EPSRC grant ref EP/X035891/1.

\section*{Data Availability Statement}

Data supporting this work will be made openly available upon journal publication.

\end{document}


\emergencystretch=3em

\title{\Large {\bf Supporting Information for}\\[4mm] ``The Zintl--Klemm Concept in the Amorphous State:\\ A Case Study of Na--P Battery Anodes''}

\author{Litong Wu}
\author{Volker L. Deringer\thanks{volker.deringer@chem.ox.ac.uk}}

\affil{Inorganic Chemistry Laboratory, Department of Chemistry, University of Oxford,\\Oxford OX1 3QR, United Kingdom}

\date{}

\maketitle

\thispagestyle{empty}

\setcounter{tocdepth}{2}
{
    \setstretch{1.4}
    \tableofcontents
}

\clearpage
\setstretch{1.5}

\section{Computational methods}

\subsection{Training dataset}

Our machine-learned interatomic potential (MLIP) model for Na--P phases was trained on a dataset developed over eight iterative training cycles. The present section outlines the creation and evolution of the training dataset.

\subsubsection{Iter-1.0} \label{Iteration 1.0}

Table~\ref{tab:iter-1.0-dataset} presents the composition of the training dataset used for the first MACE model (denoted ``Iter-1.0''). Isolated Na and P atoms, as well as 64 dimers in vacuum were created in $20 \times 20 \times 20$~\r{A}$^3$ simulation cells. The dimers were scaled to distances of 1.7 to 4.0~\r{A} for Na--Na, 1.6 to 4.0~\r{A} for Na--P, and 1.6 to 3.0 \r{A} for P--P. The distorted crystalline P, liquid P, and GAP--RSS P structures were sourced and re-labelled from the P-GAP-20 dataset.~\cite{P-GAP-20}

\vspace{3mm}

\begin{table}[h!]
\centering
\caption{Composition of the training dataset for Iter-1.0.}
\label{tab:iter-1.0-dataset}
\begin{tabular}{lcc}
\hline
\textbf{Configuration types} & \textbf{Number of structures} & \textbf{Number of atoms} \\ \hline
Isolated atoms & 2 & 2 \\
Dimers & 64 & 128 \\
\hline
Amorphous P (melt--quench) & 45 & 4,772 \\
Amorphous Na$_x$P (geometry optimization) & 405 & 56,848 \\
Distorted crystalline Na & 189 & 2,073 \\
Distorted crystalline Na$_x$P & 86 & 3,728 \\
RSS Na & 180 & 1,878 \\
\hline
Distorted crystalline P\textsuperscript & 821 & 21,382 \\
Liquid P\textsuperscript & 210 & 52,080 \\
GAP-RSS P\textsuperscript & 302 & 5,286 \\ \hline
\textbf{Total} & \textbf{2,304} & \textbf{148,177} \\ \hline
\end{tabular}
\vspace{6pt}
\raggedright

\end{table}

\textbf{Amorphous P structures.} Nine 108-atom amorphous P (a-P) structures were generated through melt--quench molecular dynamics~(MD) simulations starting from an artificial cubic lattice containing 27 uniformly spaced tetrahedral \ce{P4} molecules (P--P bond length $= 2.2$~\r{A}), set to match the density of crystalline white P. This initial structure was melted into a molecular liquid via NPT simulations at 500 K and 1 bar for 20 ps. This was followed by compression from 0.3 GPa to 1.5 GPa at 2,000 K over 100 ps to induce network liquid formation, and quenching from 2,000 K to 0.1~K at a rate of $2 \times 10^{13}$ K s$^{-1}$ at 1~bar to form a-P structures. Low-density variants were created by isotropically expanding copies of each a-P structure to densities of 2.0, 1.8, 1.6, and 1.4 g cm$^{-3}$ respectively, melting in the NVT ensemble at 1,500~K for 50 ps, and quenching to 0.1 K at $10^{13}$ K s$^{-1}$. All MD simulations were performed using LAMMPS~\cite{LAMMPS} with a timestep of 1~fs. The resulting 45 a-P structures were included in Iter-1.0 training set and used for subsequent Na insertion.

\textbf{Amorphous Na$_x$P structures.} To create amorphous Na$_x$P~(a-Na$_x$P) structures, Na atoms were randomly inserted into the previously made a-P frameworks while enforcing a hard-sphere cutoff distance of 2.2~\r{A} for Na--P interactions and 2.4~\r{A} for Na--Na interactions. Prior to Na insertion, any \ce{P4} molecules that formed in the a-P cells during NVT melt--quench simulations of the expanded structures were removed. The number of Na atoms to insert varied based on the densities of the a-P structures, with 10 and 20 Na atoms added to a-P structures at 2.0~g~cm$^{-3}$; 20, 30, and 40 at 1.8~g~cm$^{-3}$; 40, 50, and 60 at 1.6~g~cm$^{-3}$; and 60, 70, and 80 at 1.4~g~cm$^{-3}$. In addition to the a-P structures, Na atoms were also inserted into a pristine 96-atom black P cell, generating 50 different structures with 5 to 50 Na atoms added each. All Na-inserted P structures were subsequently optimized using CASTEP,~\cite{CASTEP} generating a total of 150 trajectories. Snapshots were extracted at 5\%, 30\%, and 100\% of the optimization process from each trajectory, giving a total of 450 structures. Computational details of the geometry optimization can be found in Section~\ref{optimization}. 

\textbf{Other structures.} All elemental Na and binary Na--P structures reported in the Materials Project~\cite{materials-project} were included in the dataset. Angle- and volume-distorted variants of these structures were generated by applying a random distortion in the range of ±30\% to all six lattice parameters, and 209 (Na) and 96 (Na--P) such structures were made, respectively. Additionally, 200 random Na structures were generated using the \texttt{buildcell} tool from the AIRSS package,~\cite{AIRSS-2, AIRSS} with target volumes ranging from 30 to 45~\r{A}$^3$ per atom and a minimum atomic separation of 3.0~\r{A}.

Structures in the dataset were labelled using density-functional theory (DFT) (Section~\ref{single-point}), and were split such that 10\% of each configuration type (excluding isolated atoms, dimers, and amorphous P structures) were reserved for testing, with the remainder used for training. Further details on MACE model training are provided in Section~\ref{training}.

\subsubsection{Iter-1.1 to Iter-1.4} \label{Iterations 1.1--1.4}

Using the Iter-1.0 MACE model, MD simulations were performed with the Atomic Simulation Environment (ASE) package~\cite{ase-package} to iteratively generate more a-Na$_x$P configurations. These iterative MD simulations aimed to expand both the compositional and configurational space by (i) increasing Na content up to a Na:P ratio of 3:1 and (ii) exploring the effects of elevated temperatures up to 1,500 K on a-Na$_x$P structures.

Amorphous Na$_x$P starting configurations were created using an expansion--insertion protocol similar to Section~\ref{Iteration 1.0}. Target volumes for a-P cell expansion were determined through trial and error guided by crystalline reference structures. The inserted structures were not further optimized but directly used in MACE MD simulations. In the initial simulations, a-Na$_x$P structures were heated from 300~K to 600~K over 10~ps and then equilibrated at 600~K for another 10~ps. From each MD trajectory, 20 snapshots were extracted at fixed intervals, labelled, and added into the training dataset for the next iteration of the MACE model. Once the simulations at lower temperatures stabilized, the updated MACE model was used for MD simulations at progressively higher temperatures (900~K, 1,200~K, and 1,500~K). These simulations were performed in the NVT ensemble using a Nosé--Hoover thermostat,~\cite{Nose, Hoover} with a short 0.2~fs timestep to improve stability in MD for early potential versions.

\subsubsection{Iter-2.0} \label{Iteration 2.0}

Several important changes were made in Iter-2.0:

\textbf{Removal of SEDC.} The Tkatchenko–Scheffler (TS) pairwise dispersion correction~\cite{TS-dispersion} was initially employed for optimizing Na-inserted P structures and labelling datasets used in Iter-1 models (Section~\ref{DFT}), given the known importance of dispersion interactions in accurately describing the elemental P system.\cite{dispersion-1, dispersion-2, P-GAP-20} However, as the Na content increased, it became apparent that the pairwise TS dispersion correction overestimated interactions between Na atoms, leading to unphysically short Na--Na distances and structural contraction in MD simulations. Similar issues have been reported when applying CASTEP’s semiempirical dispersion correction~(SEDC) to strongly ionic systems.~\cite{TS-challenges, Mayo-2016} To address this issue, all configurations were re-labelled with SEDCs removed.

\textbf{Removal of dimers.} Dimers were removed from the training dataset, as a test indicated that their inclusion introduced notable numerical errors and led to unstable MD simulations.

\textbf{Replacement of Na RSS structures.} Na structures obtained from non-optimized random structure search~(RSS) were replaced with GAP-driven RSS structures, following ideas and protocols described in Refs.~\citenum{GAP-RSS-1} and~\citenum{GAP-RSS-2}, and created using the \texttt{autoplex} (version 0.0.7) software.~\cite{Autoplex} In this approach, 10,000 random structures, each containing 2 to 24 Na atoms, were initially generated with per-atom volumes of 30 to 45~\r{A}$^3$ and a minimum interatomic separation of 3.0~\r{A}. Each structure was labelled using a smooth overlap of atomic positions (SOAP) descriptor,~\cite{SOAP} and the CUR algorithm~\cite{CUR} was applied to select the 100 most diverse structures. The selected structures were then subjected to single-point energy, force, and stress computations using VASP~\cite{VASP} (Section~\ref{vasp labels}) and split in a 9:1 ratio for training a Gaussian Approximation Potential (GAP).~\cite{GAP} Iterative RSS proceeded such that this GAP potential was used to minimize the enthalpy of another set of 10,000 random structures. A combination of Boltzmann-biased flat histogram sampling and leverage-score CUR selection (see Ref.~\citenum{GAP-RSS-2}) was then applied to preferentially select 100 lower-energy and structurally diverse configurations from the trajectories for the next iteration of GAP training. This process was repeated over ten iterations, and from the 1,000 structures generated, 250 were selected, labelled, and incorporated into the Iter-2.0 dataset.

Note that a few structures were lost during the re-labelling process due to incomplete CASTEP single-point computations.

\subsubsection{Iter-2.1 and Iter-2.2}

Two additional rounds of iterative MD simulations were performed to generate more a-Na$_x$P structures. The composition of the training dataset used for the final production model, Iter-2.2, is shown in Table~\ref{tab:iter-2.2-dataset}. 

\vspace{3mm}

\begin{table}[h!]
\centering
\caption{Composition of the training dataset for the final production model~(Iter-2.2).}
\begin{tabular}{lcc}
\hline
\textbf{Configuration types} & \textbf{Number of structures} & \textbf{Number of atoms} \\ \hline
Isolated atoms & 2 & 2 \\ \hline
Amorphous P (melt--quench) & 85 & 9,096 \\
Amorphous Na$_x$P (geometry optimization) & 390 & 54,290 \\
Amorphous Na$_x$P (iterative MD) & 575 & 124,522 \\
Distorted crystalline Na & 184 & 22,074 \\
Distorted crystalline Na$_x$P & 85 & 3,648 \\
GAP-RSS Na & 205 & 2,058 \\ \hline
Distorted crystalline P & 809 & 21,506 \\
Liquid P & 151 & 37,448 \\
GAP-RSS P & 297 & 5,508 \\ \hline
\textbf{Total} & \textbf{2,783} & \textbf{260,152} \\ \hline
\end{tabular}
\label{tab:iter-2.2-dataset}
\end{table}

\begin{figure}[t]
    \centering
    \includegraphics[width=\linewidth]{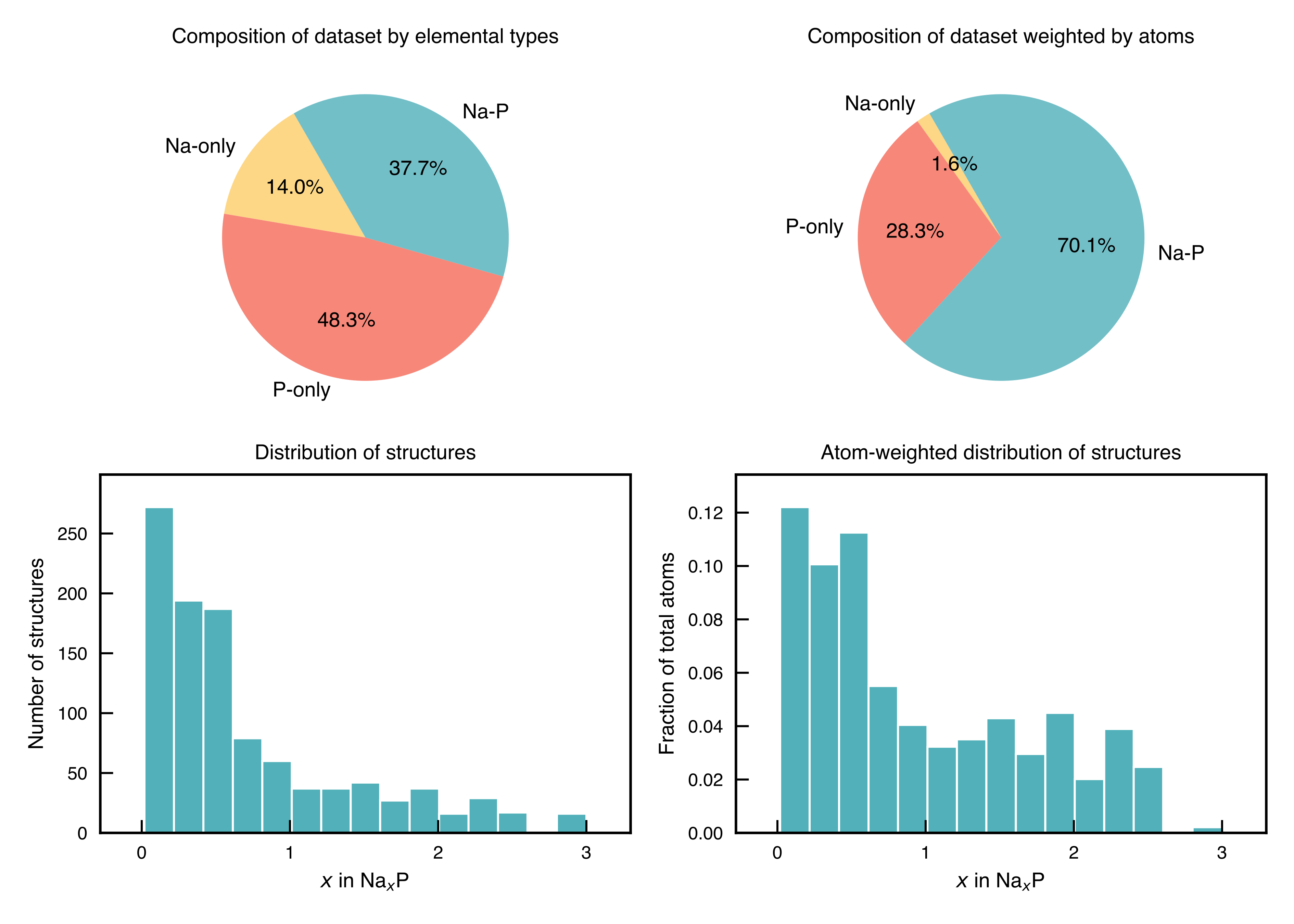}
    \caption{Structures in the Iter-2.2 training set. {\em Top:} Proportion of Na-only, P-only, and Na--P binary structures, shown by structure count (left) and atom-weighted distribution (right). {\em Bottom:} Distribution of Na$_x$P structures (left) and atom-weighted fraction of structures (right) at different Na content.}
    \label{fig:iter-2.2-dataset}
\end{figure}

The composition of the training dataset, categorized by elemental types (e.g., P-only, Na-only, and Na--P binary systems), is illustrated in the upper panel of Figure~\ref{fig:iter-2.2-dataset}. Since the a-Na$_x$P structures typically contain more atoms, the atom-weighted distribution is also included on the right. The lower panel of Figure~\ref{fig:iter-2.2-dataset} shows the distribution of Na$_x$P structures based on their Na:P ratios. The underrepresentation of configurations with Na:P ratios above 2.6 can be attributed to three main reasons: (i) no known crystalline Na--P phases exist between NaP and \ce{Na3P}; (ii) initial CASTEP geometry optimizations were only performed for a-Na$_x$P structures with $x < 1.0$; (iii) generating a-Na$_x$P structures with $x > 2.6$ through iterative MD simulations would require single-point computations of structures with over 390 atoms, which is computationally expensive at the applied level of theory. Nonetheless, subsequent studies confirmed that the MACE model remains sufficiently transferable across the compositional range.

\subsection{MACE training} \label{training}

\begin{figure}[h]
    \centering
    \includegraphics[width=\linewidth]{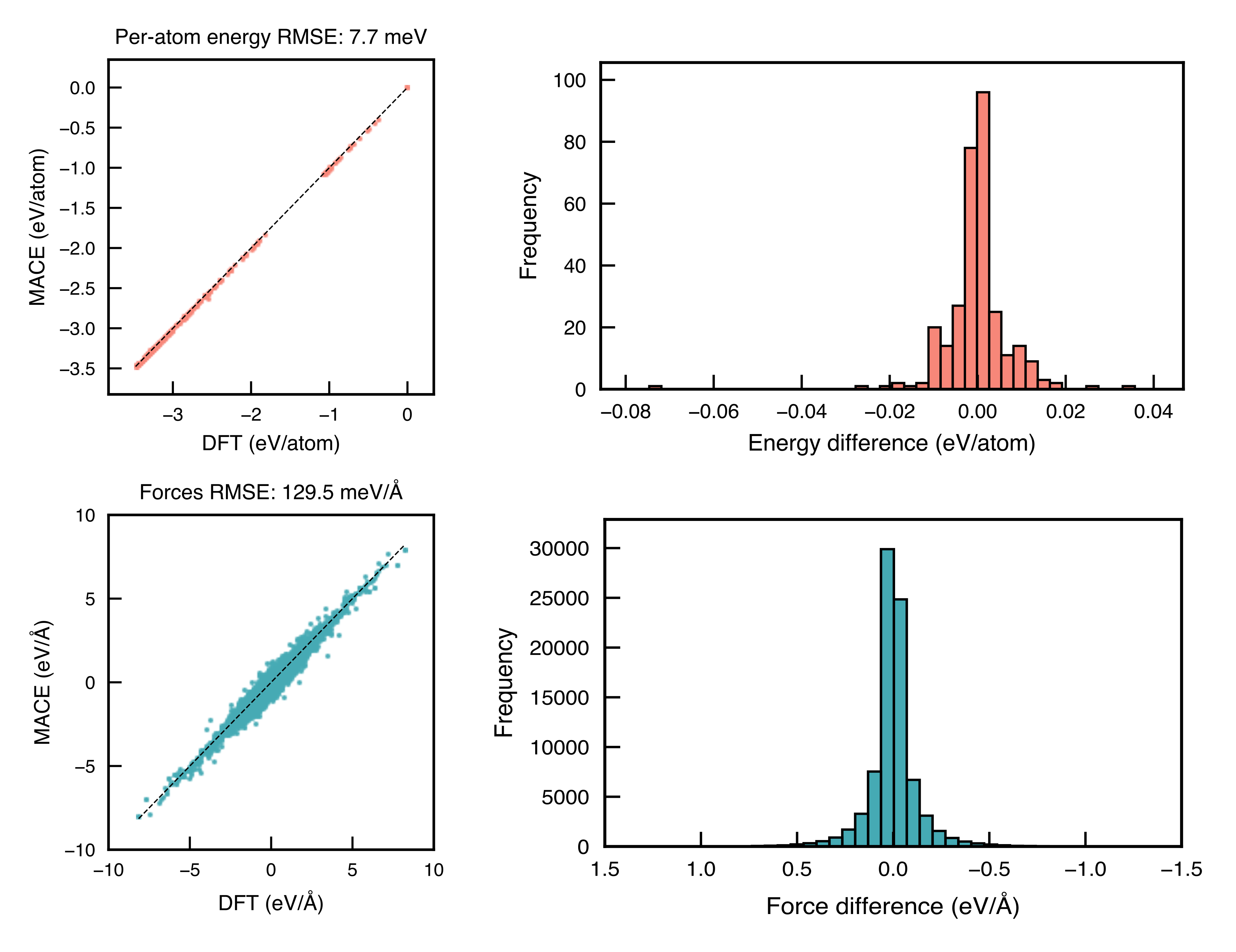}
    \caption{Numerical performance of the Iter-2.2 MACE model. {\em Left:} Parity plots comparing MACE and DFT energies (top) and forces (bottom) on the testing set. {\em Right:} Per-structure energy (top) and per-atom force (bottom) differences on the testing set.}
    \label{fig:numerical-test}
\end{figure}

MACE models were trained using a single RTX 6000 Ada Generation GPU. The dataset was split by assigning 10\% of each configuration type to the testing set, and randomly selecting 5\% of the remainder training set for validation during training. A vanilla MACE model with a model size of 128 equivariant messages and two message-passing layers was used. Each layer had a correlation order of three, an angular resolution of $l_{\text{max}} = 3$, and a cutoff radius of $r_{\text{max}} =~6.0$~\r{A}. Training was conducted using a standard weighted energy--forces loss function, with loss weights set to $\lambda_{E} = 1$, $\lambda_{F} = 100$ for the first 400 epochs, and adjusted to $\lambda_{E} = 1000$, $\lambda_{F} = 100$ for the final 350 epochs. The final MACE model achieved root-mean-square errors~(RMSE) of 7.7~meV~atom$^{-1}$ (0.74~kJ~mol$^{-1}$) for energies and 129.5~meV~\r{A}$^{-1}$ for forces on the testing set (Figure~\ref{fig:numerical-test}).

\subsection{MD simulations}

\subsubsection{Generation of low-density a-P structures}

Nine 108-atom a-P cells, along with four low-density variants of each were generated via MD simulations using LAMMPS, as detailed in Section~\ref{Iteration 1.0}.

\subsubsection{MD for iterative MACE fitting}

The iterative generation of a-Na$_x$P structures was performed using the \texttt{MACECalculator} functionality implemented in the ASE package,~\cite{ase-package} as detailed in Section~\ref{Iterations 1.1--1.4}.

\subsubsection{Production MD} \label{production md}

\textbf{Melt--quench.} Thirteen a-Na$_x$P configurations were prepared for structural and energetic analysis. A 248-atom a-P cell was isotropically expanded to densities of 1.99, 1.83, 1.47, 1.27, 1.11, 0.97, 0.87, 0.79, 0.71, 0.65, 0.60, 0.55, and 0.50~g~cm$^{-3}$ to accommodate the insertion of 31, 62, 124, 186, 248, 310, 372, 434, 496, 558, 620, 682, and 744 Na atoms, respectively. Each structure was melted at 1,200~K for 10~ps and subsequently quenched to 800~K in the NVT ensemble, using the Nosé--Hoover thermostat~\cite{Nose, Hoover} as implemented in ASE. A further quench to 5~K at 1~bar was performed in the NPT ensemble using the Berendsen scheme.~\cite{Berendsen} Two quench rates, 10$^{13}$~K~s$^{-1}$ and 10$^{12}$~K~s$^{-1}$, were tested, with their effects discussed in Section~\ref{quench-rate}. The resulting structures were optimized using the LBFGS algorithm with a force convergence criterion of $f_{\text{max}} = 0.01$~eV~\r{A}$^{-1}$.

\begin{figure}[h]
    \centering
    \includegraphics[width=0.5\linewidth]{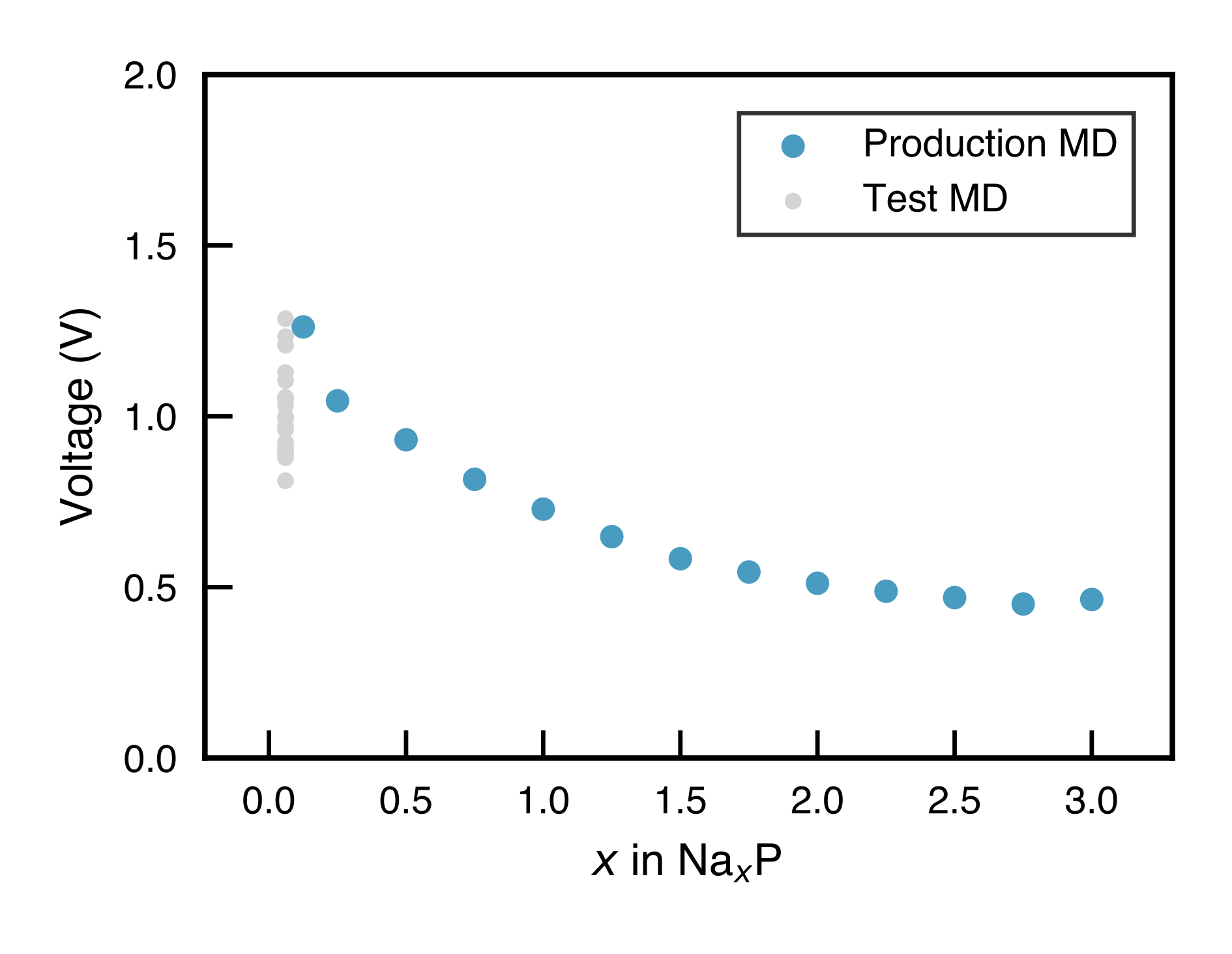}
    \caption{Effect of the MD scheme on structural energies. Grey points at $x = 0.06$ correspond to voltages calculated during initial MD tests of different melt--quench schemes and initial densities, plotted against voltages derived from the final MD scheme.}
    \label{fig:voltage}
\end{figure}

To determine the optimal MD protocol, we conducted systematic tests on a \ce{Na15P248} structure. The initial a-P density prior to Na insertion was varied between 1.8, 1.9, 2.0, and 2.1~g~cm$^{-3}$, while the switching temperature for the Berendsen NPT ensemble was tested at 500, 600, 700, and 800~K. Lower initial densities and switching temperatures increased the likelihood of \ce{P4} molecule formation. The grey points in Figure~\ref{fig:voltage} show the corresponding voltage from the test runs, highlighting the sensitivity of structural energy to simulation conditions at low Na content, where numerous insertion sites are available. However, this effect gradually diminished as Na content increased. The final MD scheme described above was selected to maximize voltage while minimizing \ce{P4} formation.

Another set of twelve 108-P a-Na$_x$P structures was generated following the same protocol, applying only the 10$^{12}$~K~s$^{-1}$ quench rate, for LOBSTER analysis (Section~\ref{Charge Calculations}). Figures~1b–d, 2a–b, and 3a present analyses based on the larger structure set, while Figures 1a and 3b–c correspond to the smaller structure set.

\textbf{High-temperature sodiation and de-sodiation.} A 256-atom black P cell was expanded by 124\% along the $x$-axis to create empty space, of which was filled with 256 Na atoms placed randomly using the same hard-sphere cutoff as described in Section~\ref{Iteration 1.0}. This cell was equilibrated at 600~K and 1~bar for 100~ps in the NPT ensemble using the Berendsen scheme, followed by a linear quench to 5~K over 60~ps. The final configuration was relaxed using the LBFGS algorithm with a force convergence criterion of $f_{\text{max}} = 0.001$~eV~\r{A}$^{-1}$. 

De-sodiation was simulated in four sequential steps, each involving the removal of 25\% of the total Na content from the relaxed sodiated structure. Atoms were selectively removed starting from those farthest from the originally P-rich region along the $x$-axis. After each removal, the structure underwent the same annealing, quenching, and optimization procedure to allow for structural reorganization before proceeding to the next extraction step.

A parallel simulation conducted at 500~K exhibited qualitatively similar structural evolution and cluster fragment statistics throughout the de-sodiation process.

\subsection{DFT computations} \label{DFT}

\subsubsection{CASTEP geometry optimization} \label{optimization}

Geometry optimizations of the initial Na-inserted a-P and black P (Section~\ref{Iteration 1.0}) were carried out using DFT implemented in CASTEP~23.1,~\cite{CASTEP} with the Perdew–Burke–Ernzerhof (PBE) functional~\cite{PBE-functional} and TS pairwise dispersion correction.~\cite{TS-dispersion} All computations were performed at the $\Gamma$ point with a plane-wave energy cutoff of 500 eV, and a SCF convergence threshold of $2 \times 10^{-6}$~eV. The optimization process used a displacement tolerance of 0.002~\r{A}, an energy tolerance of $2 \times 10^{-5}$~eV, a force tolerance of 0.05~eV~\r{A}$^{-1}$, and a stress tolerance of 0.1~GPa.

\subsubsection{CASTEP single-point computations} \label{single-point}

Single-point computations for generating training and testing data labels were computed using the PBE functional. The plane-wave cutoff energy was set to 700~eV, with an SCF convergence threshold of 10$^{-8}$~eV. The Brillouin zone was sampled using a Monkhorst--Pack grid~\cite{Monkhorst–Pack-grid} with a $k$-point spacing of $2\pi \times 0.04$~\r{A}$^{-1}$. Core electrons were described using CASTEP on-the-fly pseudopotentials.
TS pairwise dispersion correction were used for Iter-1 models but subsequently removed due to reasons described in Section~\ref{Iteration 2.0}. The energies of all reference crystalline phases considered in the analysis were computed at a consistent level of theory.

\subsubsection{VASP single-point computations} \label{vasp labels}

\textbf{GAP-RSS labelling.} Single-point computations during the GAP-driven RSS process (Section~\ref{Iteration 2.0}) were performed using the projector augmented-wave (PAW) method~\cite{PAW} in VASP~\cite{VASP} with the PBE exchange--correlation functional.~\cite{PBE-functional} A plane-wave energy cutoff of 700~eV and an electronic convergence criterion of $10^{-7}$~eV were used. The Brillouin zone was sampled using an automatically generated $k$-point mesh with a spacing of 0.2~\r{A}$^{-1}$.

\textbf{Electronic-structure computations for LOBSTER.} The PBE functional was used with a plane-wave energy cutoff of 520~eV and an SCF energy convergence threshold of $10^{-5}$~eV. Convergence tests at tighter criteria of $10^{-6}$ and $10^{-7}$~eV showed no improvement in the calculated charges within the numerical precision of the LOBSTER output. Gaussian smearing with a width of 0.05~eV was applied. An explicit $k$-point mesh with 8 points was used to sample the Brillouin zone in all cells.

\subsection{Charge computations} \label{Charge Calculations}

Atomic charges were evaluated using Löwdin population analysis, as implemented in the Local-Orbital Basis Suite Towards Electronic-Structure Reconstruction (LOBSTER).~\cite{LOBSTER, LOBSTER-charge, LOBSTER-2} The analysis was carried out on single-point wavefunctions obtained from VASP computations (Section~\ref{vasp labels}). A default minimal basis set including Na 3s and P 3s and 3p orbitals was used to project plane-wave states onto a localized atomic basis. 

\clearpage

\section{Supplementary results and discussions}
\subsection{Effects of quench rates} \label{quench-rate}

\begin{figure}[h]
    \centering
    \includegraphics[width=0.45\linewidth]{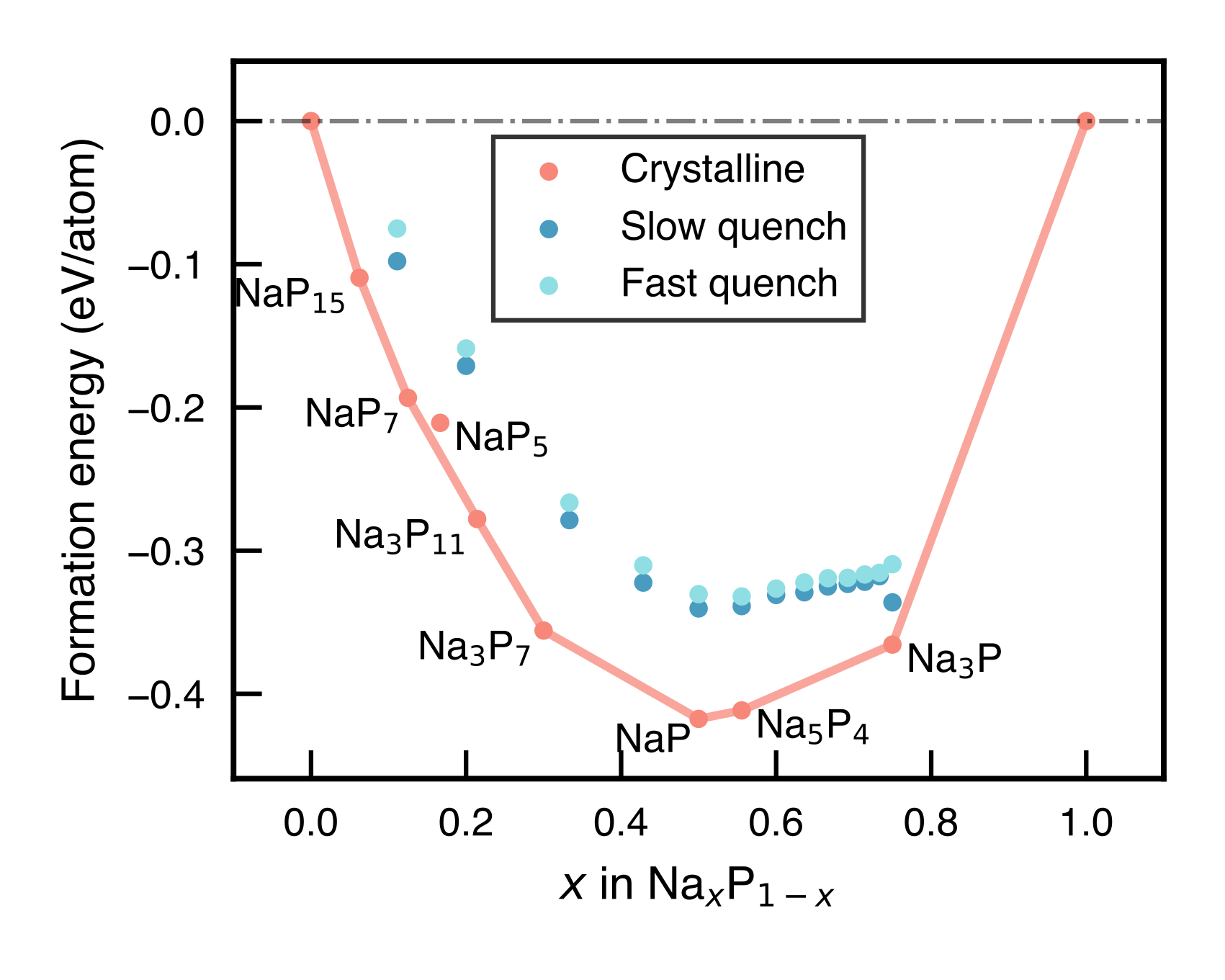}
    \caption{Convex hull including the formation energies of fast-quenched a-Na$_x$P (light blue).}
    \label{fig:convex-hull}
\end{figure}

As mentioned in Section~\ref{production md}, two quench rates --- 10$^{13}$ and 10$^{12}$~K~s$^{-1}$ --- were tested during the melt--quench generation of a-Na$_x$P structures. As shown in Figure~\ref{fig:convex-hull}, structures generated with the faster quench exhibit slightly higher normalized formation energies compared to their slow-quenched counterparts. The energy difference generally diminishes with increasing Na content, except in the case of a-\ce{Na3P} where the slow-quenched structure demonstrates markedly enhanced stabilization. 

\begin{figure}[b!]
    \centering
    \includegraphics[width=\linewidth]{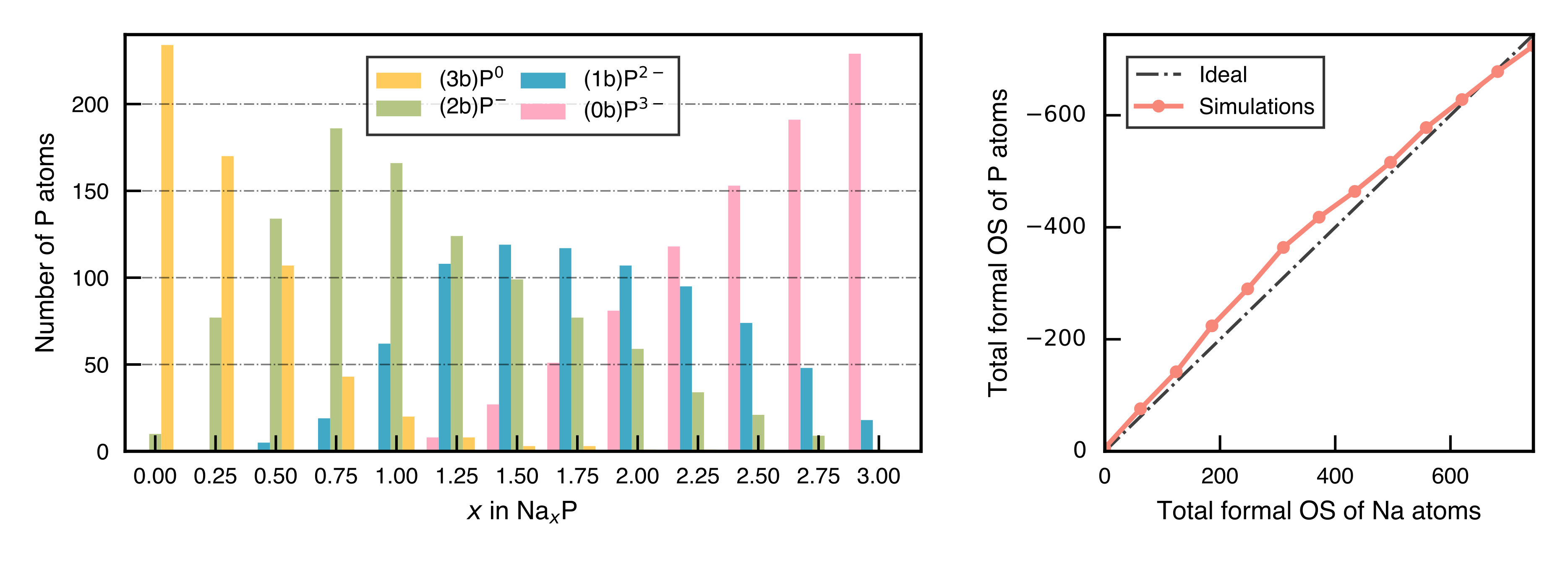}
    \caption{Coordination analysis of the fast-quenched samples. {\em Left:} Distribution of P atoms with different homoatomic coordination numbers. {\em Right:} Total formal oxidation state (OS).}
    \label{fig:zintl-bar}
\end{figure}

Despite the reduced thermodynamic stability, our fast-quenched samples also conform well with the Zintl--Klemm formalism. As illustrated in Figure~\ref{fig:zintl-bar}, the P--P coordination distributions closely resemble those shown in Figure~2a and b. However, the fast-quenched structures exhibit a modest increase in the number of (1b)P$^{2-}$ species, accompanied by a slight decrease in P atoms of other coordination types, consistent with the observation that (1b)P$^{2-}$ units are thermodynamically disfavoured.

\subsection{P--P bond lengths}

\begin{figure}[h]
    \centering
    \includegraphics[width=\linewidth]{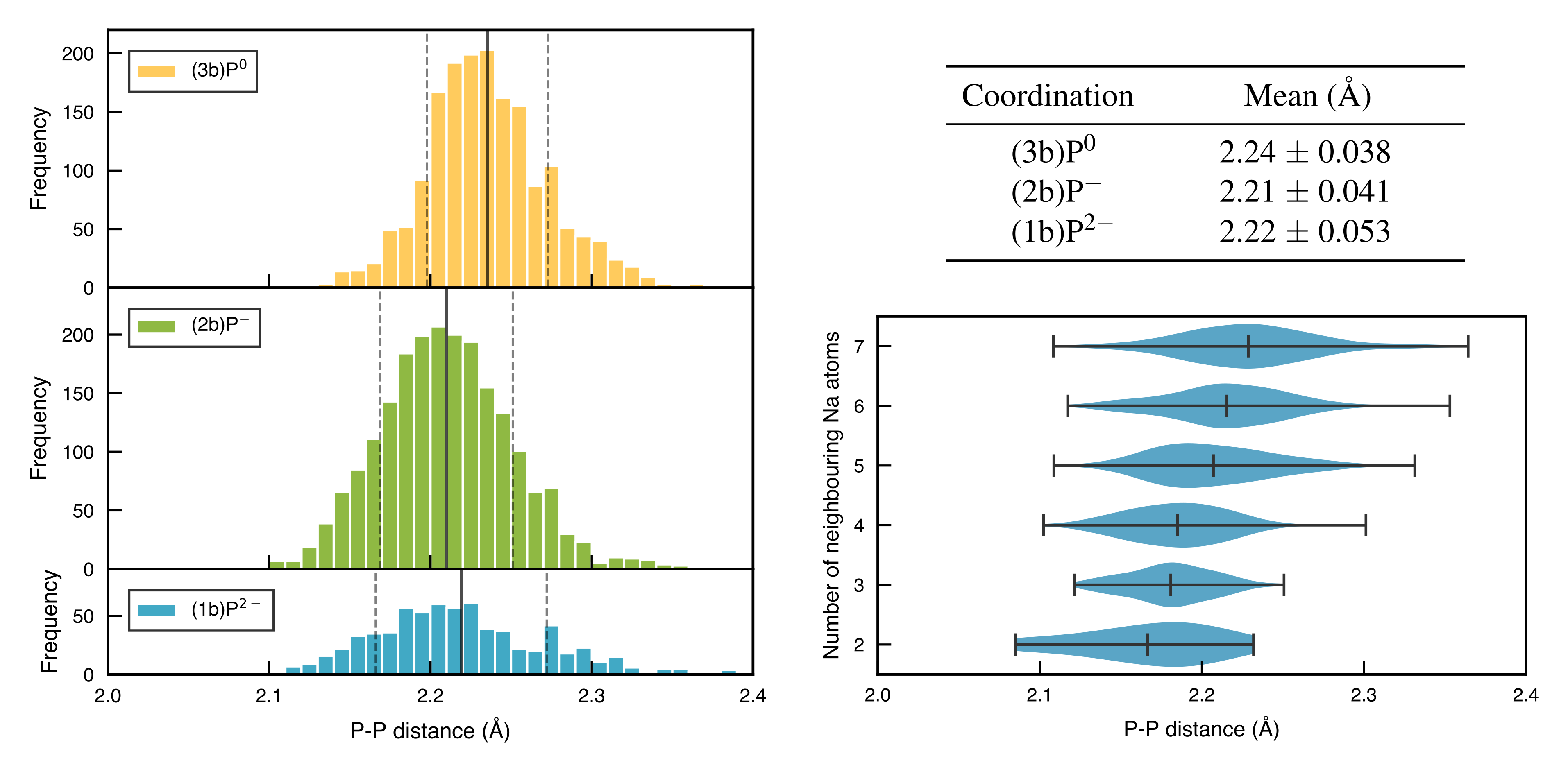}
    \caption{P--P bond length analysis. {\em Left:} P--P bond length (cutoff $= 2.4$~\r{A}) distributions, resolved by P connectivity. Solid and dotted lines indicate mean and standard deviation. Summary statistics are shown in the table at the top right. {\em Bottom right:} Correlation between the number of Na atoms within a 3.4~\r{A} cutoff and the P--P bond lengths of the (1b)P$^{2-}$ species.}
    \label{fig:P-P-bond-length}
\end{figure}

Nearest-neighbour P--P distance distributions were calculated for P with different homonuclear connectivity, and are shown in the left panel of Figure~\ref{fig:P-P-bond-length}. As expected from steric considerations, (3b)P$^{0}$ atoms exhibit, on average, longer P--P bonds than (2b)P$^{-}$ ones. However, the (1b)P$^{2-}$ ions deviate from the steric expectation, exhibiting both a larger average than (2b)P$^{-}$, and a larger spread in bond lengths, as indicated by their higher standard deviation.

To better understand this anomaly, we analyzed the influence of neighbouring Na atoms (within a 3.4~\r{A} cutoff) on the P--P bond lengths of the (1b)P$^{2-}$ species. The violin plot on the bottom right of Figure~\ref{fig:P-P-bond-length} shows that, for the (1b)P$^{2-}$ species, an increasing number of nearby Na atoms correlates with larger average and maximum nearest-neighbour P--P distances. This bond lengthening effect may arise from electron transfer from adjacent Na atoms, which could weaken the P--P interactions through increased electron density localization on the P sites.

\subsection{Atomic and environmental energies of Na atoms}

\begin{figure}[h]
    \centering
    \includegraphics[width=\linewidth]{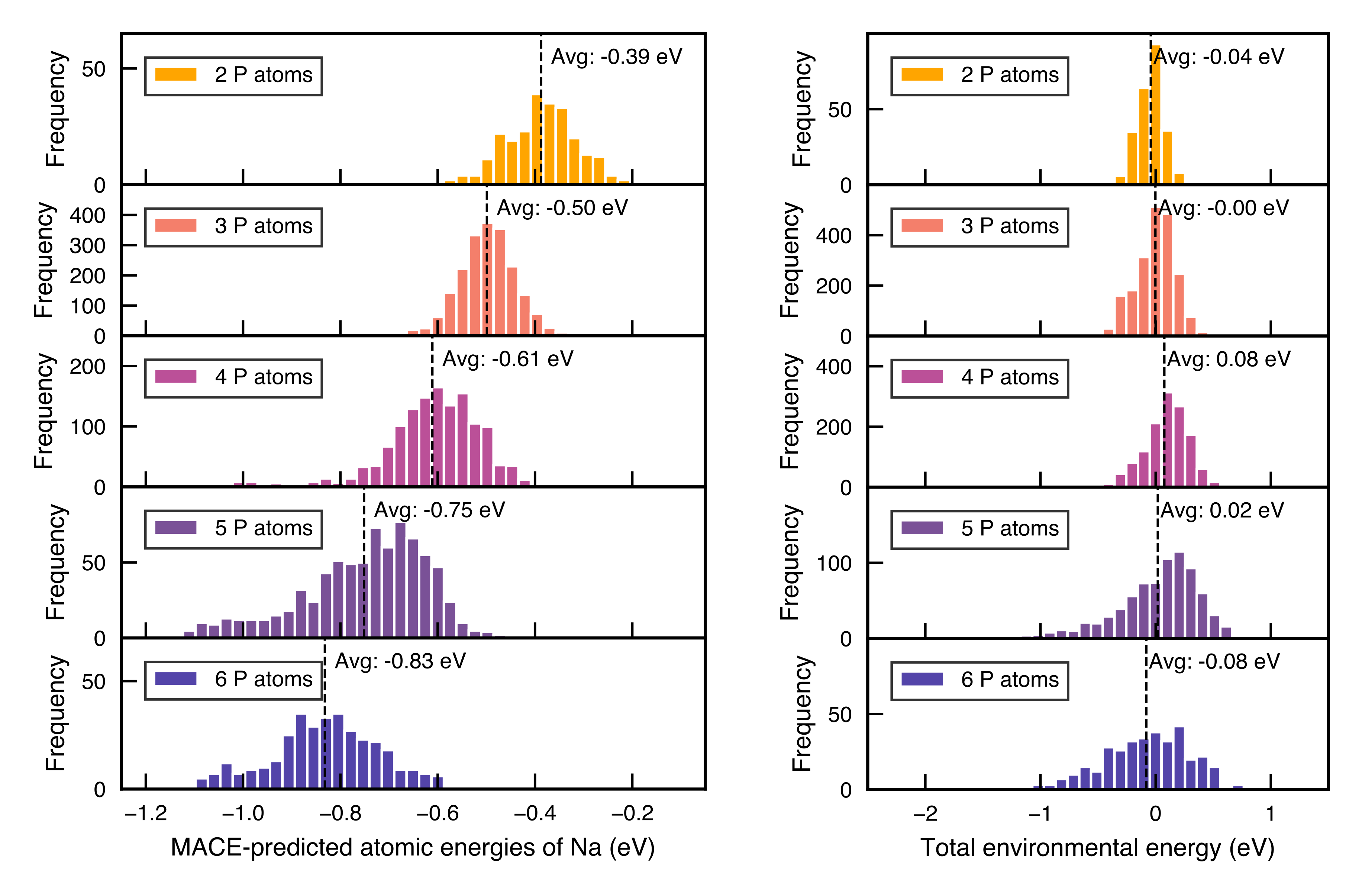}
    \caption{Local energy analysis of Na in a-Na$_x$P. {\em Left:} MACE atomic energies of Na atoms, grouped by the number of neighbouring P atoms within a 3.4~\r{A} cutoff. Energies are referenced to bulk body-centered cubic Na. {\em Right:} Local environmental energies of Na atoms, calculated as the sum of the Na atom energy and energies of neighbouring P atoms within the same cutoff.}
    \label{fig:Na-energies}
\end{figure}

The MACE-predicted atomic energies of Na atoms in a-Na$_x$P, categorized by the number of neighbouring P atoms within a 3.4~\r{A} cutoff, are shown in the left panel of Figure~\ref{fig:Na-energies}. A clear stabilization trend is observed as the number of neighbouring P atoms increases, possibly implying stronger local ionic interactions in more P-rich environments.

The right panel of Figure~\ref{fig:Na-energies} shows the corresponding local environmental energies, defined as the sum of the central Na atom energy and energies of neighbouring P atoms within the same cutoff distance. Interestingly, the average environmental energy across different Na coordination environments remains close to 0~eV, suggesting that the local environments around Na atoms are energetically compensated for by nearby P atoms. The interpretation of this energetic balance warrants further investigation.

\subsection{Atomic energies versus charges}

\begin{figure}[h]
    \centering
    \includegraphics[width=0.9\linewidth]{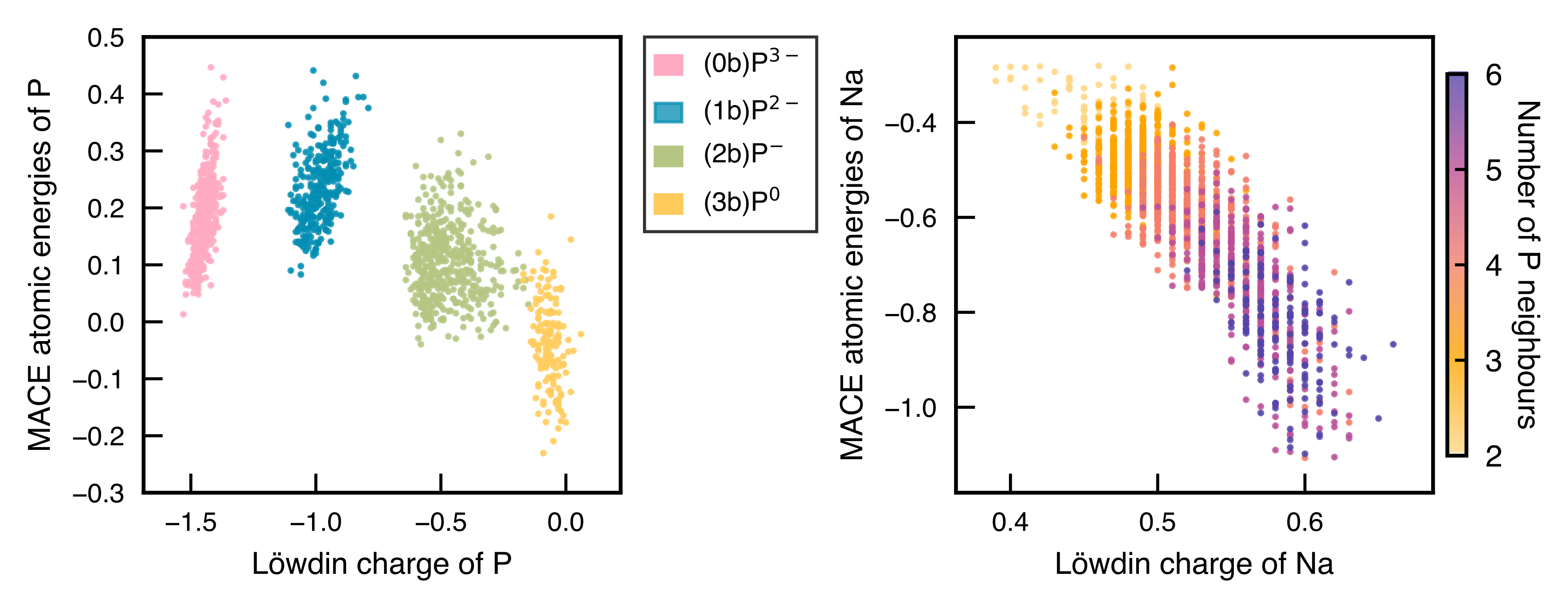}
    \caption{MACE atomic energies against Löwdin charges for individual P (left) and Na (right) atoms in a-Na$_x$P structures. P atoms are colour-coded based on local homonuclear connectivity; Na atoms are colour-coded according to the number of neighbouring P atoms within a 3.4~\r{A} cutoff.}
    \label{fig:energy-charge-scatter}
\end{figure}

To better visualize the correlation between atomic energy and charge, the MACE energies of individual atoms are plotted against their respective charges and shown in Figure~\ref{fig:energy-charge-scatter}. The distribution of P atom energies mirrors the trends observed in Figure~3, with (1b)P$^{2-}$ species exhibiting the highest average energy among all P coordination classes. Distinct charge separation is evident across different P species, where decreasing homoatomic connectivity correlates with increasingly negative Löwdin charges.

In contrast, Na atoms have a more continuous relationship between energy and charge, where Na atoms with a greater number of neighbouring P atoms (purple) tend to exhibit more positive Löwdin charges and lower MACE atomic energies. This trend highlights the role of local Na--P coordination in stabilizing Na atoms: the more an Na atom is embedded within the P network, the greater the charge transfer. The resulting stronger ionic interactions are then reflected in the lower atomic energy. 

\subsection{Ring and cluster fragment statistics during de-sodiation}

\begin{figure}[h]
    \centering
    \includegraphics[width=0.7\linewidth]{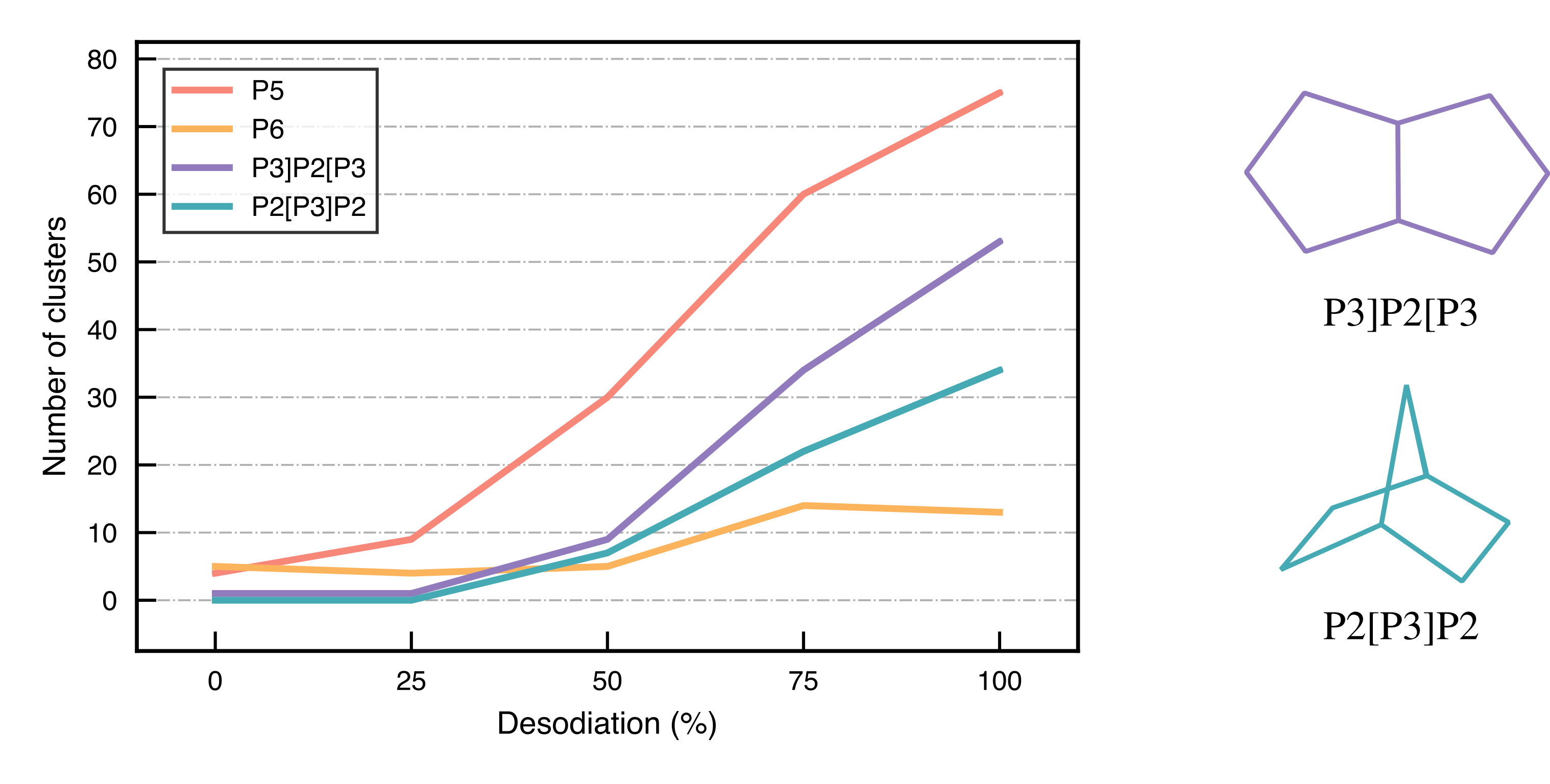}
    \caption{Number of rings and local cluster fragments after 25\%, 50\%, 75\%, and 100\% de-sodiation. Illustrations of P3]P2[P3 and P2[P3]P2 cluster fragments (cf.\ Ref.~\citenum{Bocker-1995}) are shown on the right.}
    \label{fig:cluster-counting}
\end{figure}

Ring and cluster analysis during de-sodiation simulations was performed using a custom Python script to track local structural evolution. As shown in Figure~\ref{fig:cluster-counting}, progressive Na extraction led to a steady increase in the number of primitive five-membered rings and local cluster fragments, specifically P3]P2[P3 and P2[P3]P2 (illustrated on the right). In contrast, the number of six-membered rings, characteristic of crystalline black P, increased only marginally. This asymmetry in the recovery of structural motifs suggests that the structural disorder introduced during sodiation is largely irreversible, with the de-sodiated structure tending toward locally disordered amorphous motifs.

\subsection{Effects of many-body dispersion corrections}

\begin{figure}[b]
    \centering
    \includegraphics[width=0.7\linewidth]{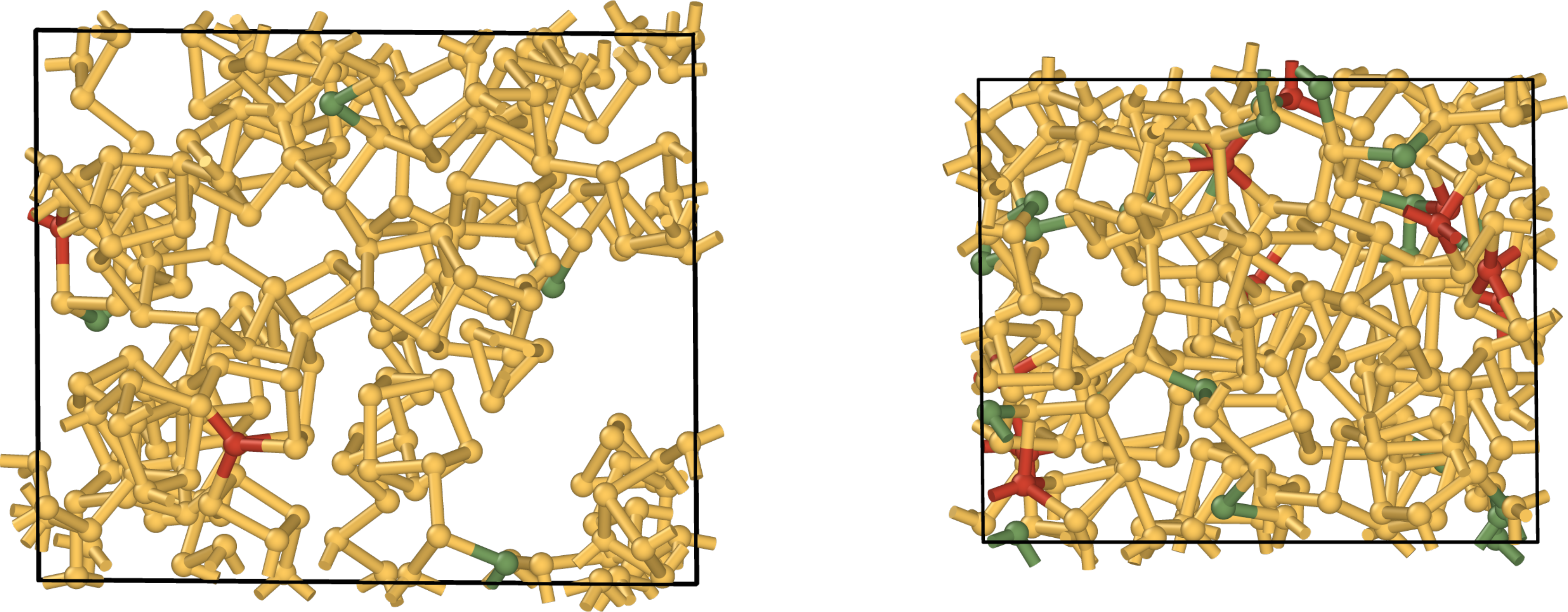}
    \caption{Final structures of the fully de-sodiated cell after relaxation using the MACE model (left) and the P-GAP-20 model from Ref.~\citenum{P-GAP-20} (right). Cell sizes are shown to scale.}
    \label{fig:MBD-effect}
\end{figure}

As discussed in Section~\ref{Iteration 2.0}, dispersion corrections were excluded from Iter-2.0 onward, as their influence became increasingly unclear in a dataset dominated by ionic structures. Nevertheless, previous studies have emphasized the importance of dispersion corrections in accurately describing elemental P systems.~\cite{dispersion-1, dispersion-2, P-GAP-20} Since the fully de-sodiated structure is an elemental P system, we evaluated the potential impact of dispersion corrections on those structures by repeating the same simulation protocol using the P-GAP-20 model. This model had been trained at the PBE+MBD level of theory and incorporates a 1/$r^6$-dependent baseline to account for long-range interactions.~\cite{P-GAP-20}

After holding at 600 K and 1 bar for 100 ps, and quenching at 10$^{13}$~K~s$^{-1}$ (Section~\ref{production md}), the final structure obtained with the P-GAP-20 model was noticeably more contracted than that obtained with the Na--P MACE model of the present work, consistent with the absence of long-range dispersion corrections in the training labels of the latter.

Despite the large difference in final density --- 1.47 versus 2.46 g~cm$^{-3}$ --- both models produced similar local structural features, as reflected in the comparable fragment counts in Table~\ref{tab:MBD-effect}. This suggests that the MACE model remains robust in reproducing local bonding environments even at expanded volumes. We therefore attribute the expansion observed in the MACE-relaxed structure to limitations in the training labels rather than to a physically meaningful effect.

\vspace{3mm}

\begin{table}[h!]
\centering
\captionsetup{width=0.9\textwidth}
\caption{Comparison of density and fragment counts of fully de-sodiated structures obtained from the Na--P MACE model of the present work and the elemental P-GAP-20 model.~\cite{P-GAP-20}}
\begin{tabular}{>{\centering\arraybackslash}m{2.5cm}
                >{\centering\arraybackslash}m{3cm}
                >{\centering\arraybackslash}m{1.4cm}
                >{\centering\arraybackslash}m{1.4cm}
                >{\centering\arraybackslash}m{1.7cm}
                >{\centering\arraybackslash}m{1.7cm}}
    \Xhline{2\arrayrulewidth}
    \multirow{2}{*}{\textbf{ML models}} & \multirow{2}{*}{\textbf{Density (g cm$^\mathbf{-3}$)}} & \multicolumn{4}{c}{\textbf{Fragment counts}} \\
    \cline{3-6}
    & & \textbf{P5} & \textbf{P6} & \textbf{P3]P2[P3} & \textbf{P2[P3]P2} \\
    \hline
    Na--P MACE & 1.47 & 75 & 13 & 53 & 34 \\
    \hline
    P-GAP-20 & 2.46 & 75 & 20 & 52 & 39 \\
    \Xhline{2\arrayrulewidth}
\end{tabular}
\label{tab:MBD-effect}
\end{table}

\subsection{Structural fragments in related crystalline phases}

\begin{figure}[h]
    \centering
    \includegraphics[width=0.7\linewidth]{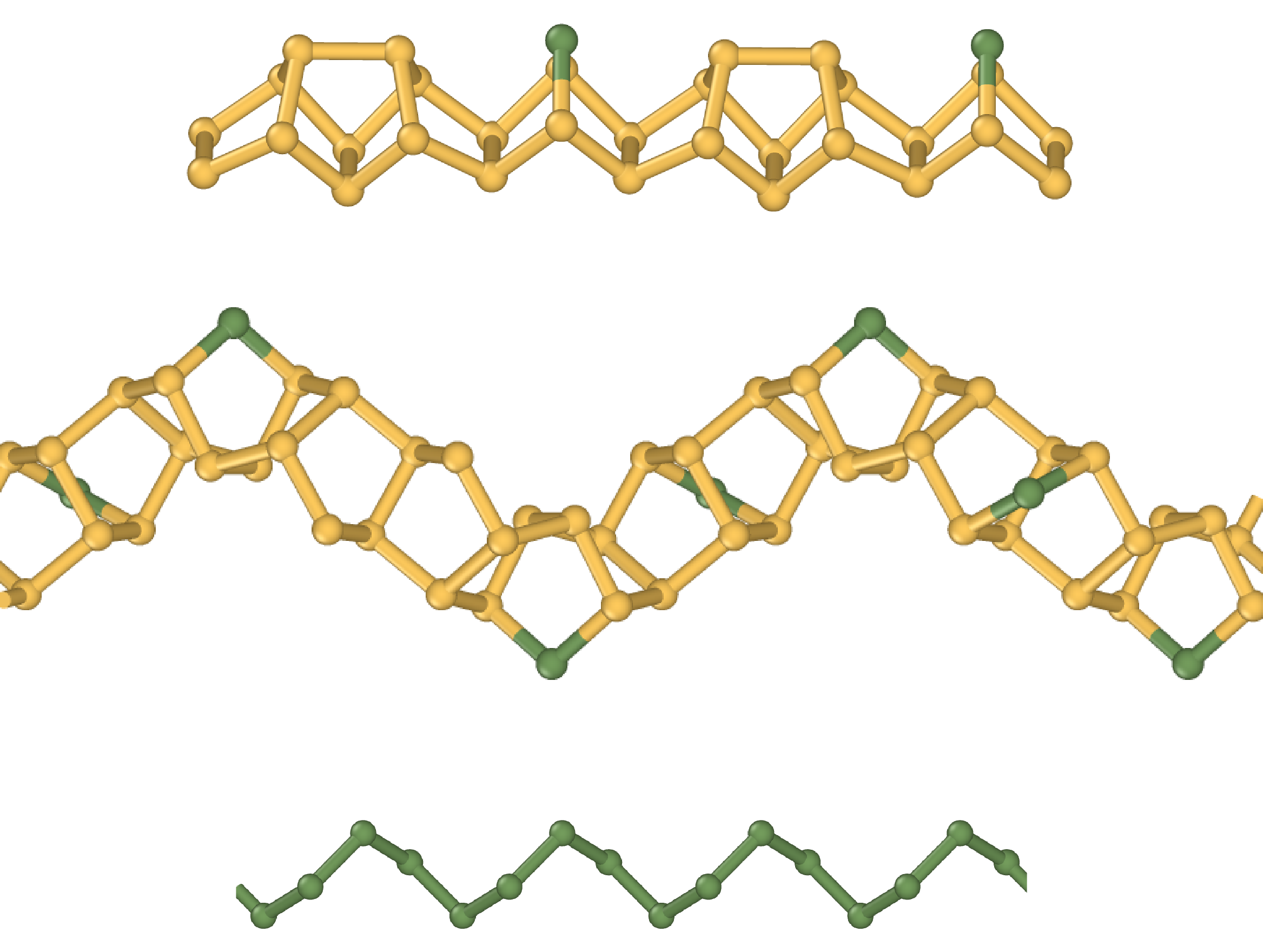}
    \caption{P structural framework in crystalline \ch{NaP15} (top), \ch{NaP7} (middle) and NaP (bottom), shown for comparison with Figure~2c. In \ch{NaP15}, \ch{^1_{$\infty$}(P15^-)} tubes form through the polymerization of alternating \ch{P8} (similar to the \ch{As4S4} molecular structure) and \ch{P7^-} (a norbornane derivative) units. In \ch{NaP7}, \ch{^1_{$\infty$}(P7^-)} helical tubes form by connecting norbornane-like \ch{P7^-} clusters via some of the base and the bridging P atoms. The anionic partial framework in NaP consists of \ch{^1_{$\infty$}(P^-)} helical chains.} 
    \label{fig:crystal-fragments}
\end{figure}
 
\clearpage

\setstretch{1.1}

\section{Supplementary references}